\documentclass[traditabstract]{aa}

\usepackage{graphicx}
\usepackage{amsmath}
\usepackage{natbib}
\bibpunct{(}{)}{;}{a}{}{,} 

\newcommand{\dd}{\,\mathrm{d}}

\begin{document}

\title{Influence of baryonic physics in galaxy simulations:}

\subtitle{a semi-analytic treatment of the molecular component}

\author{A. Halle        \inst{1}
        \and
       F. Combes        \inst{1}
}

\offprints{A. Halle}

\institute{Observatoire de Paris, LERMA, 61 Av. de
               l'Observatoire, 75014 Paris, France\\
                \email{anaelle.halle@obspm.fr}, \email{francoise.combes@obspm.fr}   
                }

\date{}

\abstract{ Recent work in galaxy formation has enlightened the important role of baryon physics, to solve the main problems encountered by the standard theory at the galactic scale, such as the galaxy stellar mass functions, or the missing satellites problem. The present work aims at investigating in particular the role of the cold and dense molecular phase, which could play a role of gas reservoir in the outer galaxy discs, with low star formation efficiency. Through TreeSPH simulations, implementing the cooling to low temperatures, and the inclusion of the molecular hydrogen component, several feedback efficiencies are studied, and results on the gas morphology and star formation are obtained. It is shown that molecular hydrogen allows some slow star formation (with gas depletion times of $\sim 5$~Gyr) to occur in the outer parts of the discs. This dense and quiescent phase might be a way to store a significant fraction of dark baryons, in a relatively long time-scale, in the complete baryonic cycle, connecting the galaxy discs to hot gaseous haloes and to the cosmic filaments.

}

\maketitle

\section{Introduction} 

Numerical simulations of galaxy formation have now reached a high degree of sophistication, and include increasingly detailed physics, from star formation  rates treated with sub-grid recipes based on the Kennicutt-Schmidt law (KS), to more direct processes triggered by Jeans instabilities \citep{stinson06,hopkins11}, feedback that is treated either through supernovae heating, or kinetic impulse, momentum-driven flows due to stellar winds \citep{sales10,ostriker11}, and in some cases, multiphase gas \citep{maio07,gnedin09}. However, big unknowns remain as free parameters in the process: first the nature of dark matter, and its behaviour, leading to well-identified problems at galaxy scales, such as predicted cuspy profiles (while only cores are observed), or predicting a large number of dark satellites around each Milky-Way type galaxies.  Second, a large unknown is also the nature and location of the missing baryons: observationally only less than half of the baryons have been identified \citep[e.g.][]{fuku04}, and this missing baryonic component is certainly gaseous, given the results of microlensing \citep[e.g.][]{wyr09}. This gas might be hot (of the order of millions of degrees) or cold, and most of it must be located in the intergalactic space, not to overpredict rotation curves.

A significant fraction of dark baryons should also exist in galaxies, and could reside under the form of cold gas, dense enough to be in the molecular phase \citep[e.g.][]{Pfe94,gren05,bour07,langer10}. A reservoir of cold gas may help to moderate star formation, and explain almost stationary star formation histories in late-type galaxy discs as observed today \citep[e.g.][]{wyse09}. When the thermal evolution of gas is studied during galaxy formation, an immediate conclusion is that  almost all the baryons should have cooled and condensed in dark matter potential wells, where they are supposed to form stars, which are not seen. This overcooling problem can be solved through re-heating the gas, even before its inflow in the dark matter bound structures \citep{blan92,dave01}. However the star formation efficiency might be lower in  globally low gas density environments, such as the outer parts of galaxies, and  cold gas reservoirs are another solution to explore. The time-scale spent by the gas in the cold phase is largely unknown, and this could change the baryon cycle. Cosmological simulations are unable to derive the abundance of cold gas, by lack of spatial resolution, leading to reduce density dynamics, underestimating the cooling (that increases with density) locally.

In the recent years, the cold mode gas accretion to assemble mass onto galaxies has been  revived versus the hot mode gas accretion. It has been realised that only part of the gas was heated to the virial temperature of a structure before cooling and inflowing, and most of the gas could be accreted cold \citep{keres05}. The cold gas accretion is dominating for small haloes, lower than $10^{12}$ M$_\odot$, and is inflowing along the cosmic filaments, while for massive systems, the hot quasi-spherical mode dominates. There is also a redshift dependence on the importance of the two modes, the cold mode being more dominant at high redshift ($z$ larger than 2). \citet{birn03} and \citet{dekel06} found similar conclusions, using one-dimensional simulations and analytic arguments. The limit between dominating cold and hot modes corresponds to a baryonic mass close to the scale separating the red sequence and the blue cloud in galaxy populations \citep{kauf03,bald04}, and the origin of bimodality in observed galaxies has been connected to these two different theoretical assembly modes. Even in massive structures, cosmological simulations have shown that cold filaments of gas may penetrate their hot gaseous halo \citep{birn07,dekel09}.

A further well-known problem in the baryon cycle and galaxy formation is to reproduce correctly the galaxy stellar mass function. If the large amount of predicted dwarf galaxies are suppressed by supernovae feedback, assumed to expel gas out of their small potential wells, gas is still accreted in massive systems, and forms constantly growing, blue star-forming galaxies at the massive end of the mass function. Solutions have been proposed through AGN feedback (either quasar mode through heating and outflows, or radio mode through powerful radio jets), with limited success however. In particular, the gas expulsion is more efficient again in low-mass galaxies. Besides, even when ejected with a speed larger than the escape velocity, gas is re-accreted onto galaxies, being braked not only through gravity but also by hydrodynamical interactions with the halo gas. The re-accretion has been called halo fountain by \citet{Oppen08}. According to the environment, gas outflowing from low or intermediate mass haloes can be re-accreted by more massive companion galaxies, a process which has been dubbed the intergalactic fountain \citep{keres09}. Globally, the frequent re-accretion of the ejected gas limits the efficiency of the AGN feedback, maintaining some overcooling problem, and keeping massive galaxies active in star formation until late times, with no definite quenching. It is possible to introduce in the baryon cycle a cold gas reservoir, with a sufficiently long time-scale in this phase to account for the observed properties of galaxies. In the present paper, we want to explore this possibility, beginning by isolated galaxy simulations, and considering in a companion paper galaxies in their environment. We are taking into account a cold dense molecular phase in the gas component, which can cool at a temperature much lower than the 10$^4$K usually assumed for ``cold gas'' in cosmological simulations. According to the baryon physics adopted, and in particular the amount of feedback, we will examine the stability of the galaxy discs, and the efficiency of star formation, in order to derive the order of magnitude of the time-scale spent by the gas in this cold component reservoir.

\smallskip

Thanks to the always increasing computing capacities, there has been in recent years a growing number of simulations taking into account the cold phase. These works come from two different  domains, with different goals:
\begin{enumerate}

\item Cosmologically oriented work, tending to higher spatial resolution, of the order of 100~pc; their main interest is to determine whether the small scale structure impacts the large scale tranfer of angular momentum, the stability of discs, formation of bulges, concentration of the galaxy, transformation of galaxies from late to early type, whether feedback can destroy dark matter cusps, etc.. \citep{maio07,gnedin09,schaye10,mura10,bour10,hopkins11};

\item Star formation oriented work, tending to enlarge the field of view from molecular clouds to the whole galaxy structure. These simulations are in general multiphase, with the goal to answer questions about the cloud formation, the star formation efficiency The galaxy models are often not self-consistent, with no transfer of angular momentum from gas to stars or dark matter halo. \citep{slyz05,tasker06, tasker08, dobbs08, dobbs11, shetty08, wada11}. 

\end{enumerate}
Some works are at the transition between the two scales, and when the simulations are unable to resolve the cloud fragmentation scale, have tried to represent the multi-phases of the gas using sticky particles, moving ballistically in the potential, and able to collide and coagulate to form a whole mass spectrum of clouds \citep[e.g.][]{seme02,booth07, revaz09}.

In the present work, our goal is to deal with the multiphase of the gas in a self-consistent way, allowing the gas to cool down to 100~K, the average temperature of the quiescent dense molecular phase, and to follow its interaction with the other phases through cooling/heating, star formation and feedback, through a Tree-SPH code. The H$_2$ molecule formation will be simplified by a recipe related to the density, metallicity and self-shielding of the gas. The star formation recipe is based on the Kennicutt-Schmidt law, and several variations will be explored,  depending on the volume or surface density of the gas, to take into account the large variations of star formation efficiency with the environment. Various feedback physics will be tested, and both the stability of the disc will be studied, with respect to bar and spiral formation, and the morphological structure of the gas will be quantified.

\smallskip

The numerical techniques and the initial conditions of the simulations carried out are described in \S \ref{simu}, together with the details of the baryonic physics adopted to deal with cold molecular hydrogen, star formation and feedback. \S \ref{resu} describes the results of the simulations. The influence of the feedback and inclusion of molecular hydrogen cooling is detailed. Our conclusions are drawn in \S \ref{conclu}.

\section{Numerical techniques} 
\label{simu}

We use the Gadget-2 code \citep{springelg2} that computes gravitational and hydrodynamical forces. Gravity is computed by a Tree algorithm with a Barnes Hut opening angle $\theta$ of $0.7$, taking only the monopole moments of the gravitational field into account for computational reasons (see \citet{springelg2} for details). We take the same constant softening length $\epsilon$ for all particle types (stars, gas and dark matter). Hydrodynamics is treated with a Smooth Particle Hydrodynamics (SPH) algorithm, with an individual smoothing length $h$ computed such that a constant mass is contained in a sphere of radius $h$, and that is allowed to be as low as 0.1 $\epsilon$.


We add baryonic physics to this code, as described in this section.  

\subsection{Cooling} 

Gas in the interstellar medium (ISM) loses internal energy by radiation due to physical processes occurring at microscopic scales. This cooling can be encapsulated in a cooling function representing the corresponding volume rate of energy loss, and depending on the chemical composition of the gas, its density and temperature. In simulations of galaxies, cooling is often considered only down to a temperature of $\sim$ 10$^4$~K, as the most efficient cooling processes are due to ionised hydrogen and helium, elements that are almost only present in the atomic form below 10$^4$~K. This usually implies the coldest dense gas is at an equilibrium temperature of about 10$^4$~K in these simulations, as dense gas experiences heating pressure forces and shocks if it reaches lower temperatures, and the gas is prevented from collapsing further. We include cooling by metals, molecular hydrogen and HD down to 100~K, a more realistic temperature floor for the ISM.

In dense media, cooling by metals might dominate at solar metallicity. However, metals are abundant only in the central parts of galaxies, and radial abundance gradients are observed in giant spiral galaxies, with an exponential decline of about 0.6 dex in 10 kpc on average \citep[e.g.][]{henry95,vanzee98}. Therefore the outer parts of galaxies are reminiscent of the primordial galaxies, with low metal abundance. Cooling through the H$_2$ and HD molecules is then important, and could change the physics of star formation in these regions. Ultraviolet observations by GALEX have shown that star formation can be active at large radii, much farther than the optical radius R$_{25}$, in regions where H$_{\alpha}$ observations are not able to reveal moderate-age populations of stars. UV-bright discs extend up to $3-4$ times the optical radius in about 30 $\%$ of spiral galaxies \citep{thilker05,gildepaz05,gildepaz07}. The presence of molecular hydrogen and star formation in outermost discs of spirals is of prime importance to study cold gas accretion, which is considered one key factor in galaxy evolution \citep[e.g.][]{keres05,dekel09}.

\subsubsection{H, He and metals cooling above $10^4$~K}

We use the cooling functions computed by \citet{suther93} for a plasma in collisional ionisation equilibrium above $10^4$ K. These cooling functions include cooling due to ionised H, He and metals and are tabulated for different metallicities. 

\subsubsection{Metal-line cooling below $10^4$~K}

Following \citet{maio07}, we compute cooling functions of a few metals below $10^4$~K: CII, OI, SiII and FeII, as they are the most abundant heavy elements released by stars in the ISM and are thus the main ingredients for cooling, considering their collisions with H atoms and electrons. For ``low'' densities obtained in galaxy simulations, the populations of energy levels needed to compute the cooling rates differ from the Boltzmann populations of the Local Thermal Equilibrium (LTE). The populations here depend on the abundances of species that can collide with the metal and allow it to change of energy level, which makes the rate of energy loss per metal depend on these abundances (contrarily to the LTE case for which this rate depends only on temperature). The resolution of the equations governing the populations is included in the code, using the quantum data given in \citet{maio07}, and the cooling functions are computed using the obtained populations and the transition probabilities, as in \citet{maio07}. 

We choose the same abundances as in \citet{suther93} (solar abundances and primordial ratios) for the metals we consider here and a low electronic fraction of ($\frac{n_{\mathrm{e^-}}}{n_{\mathrm{H}}}=10^{-5}$), assuming the cold gas is quasi neutral.

\begin{figure}[!h]
\centering
\resizebox{\hsize}{!}{\includegraphics{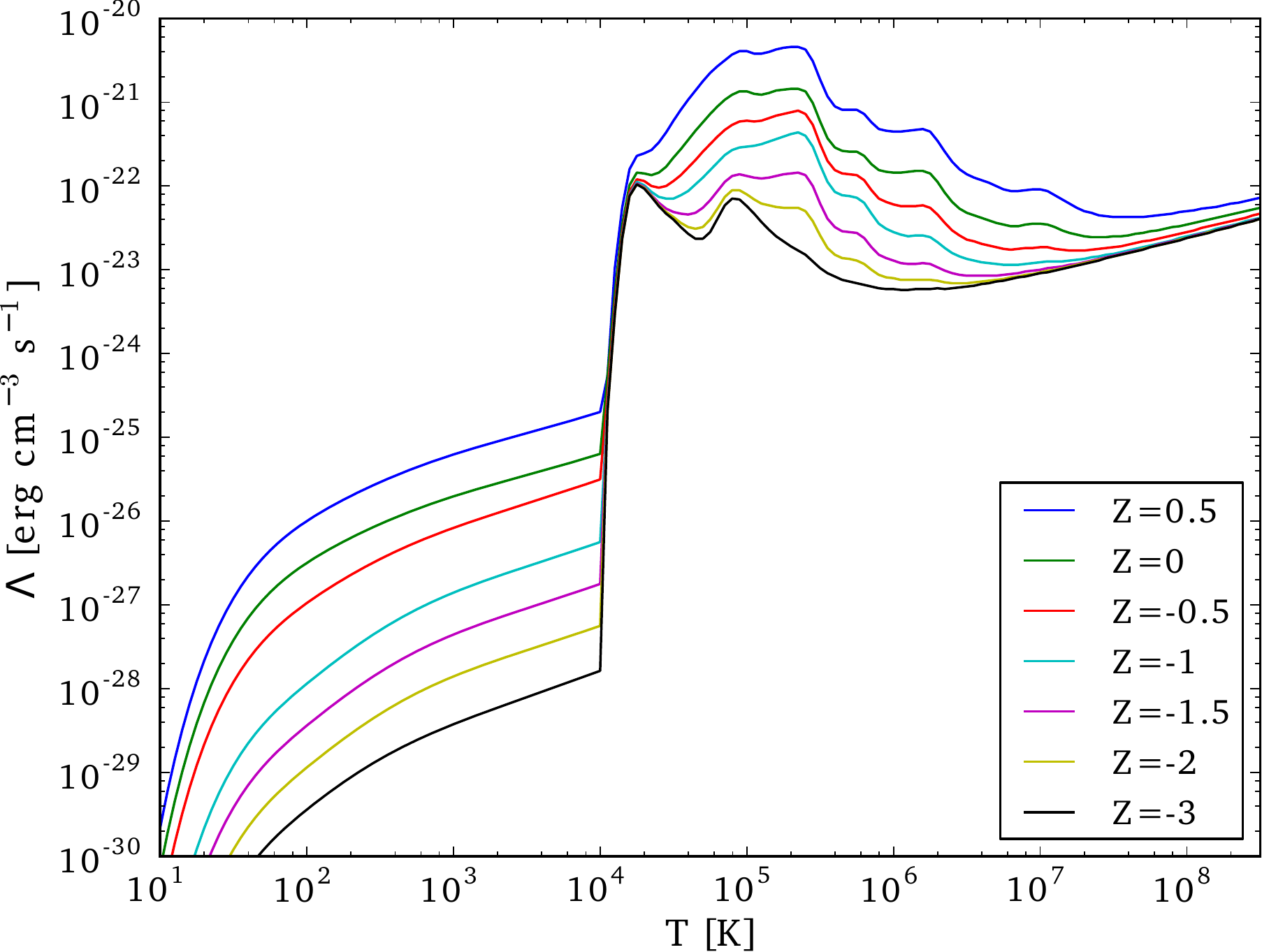}}
\caption{Volume cooling rates for n$_H$=1~cm$^{-3}$ , following \citet{suther93} for high temperatures ($T$ $>$ 10$^4$ K), and \citet{maio07} for low temperatures ($T$ $<$ 10$^4$ K), as a function of temperature and metallicity $Z=[\mathrm{Fe/H}]$ ($Z=0$ is solar). In the code, we use exact density-dependent cooling functions for the low temperature part, cf Sec. 2.1.}
\label{cool-fig}
\end{figure}

The cooling curves for n$_{\mathrm{H}}=1 \, \mathrm{cm}^{-3}$ and different metallicities $Z=[\mathrm{Fe/H}]$ are plotted in Figure~\ref{cool-fig}.

\subsubsection{H$_2$ cooling}

We are interested in studying the influence of molecular hydrogen cooling on galaxies, especially in regions where metals are not abundant. We tabulate the LTE cooling function as a function of temperature from the Boltzmann energy levels populations and transition probabilities. For low densities for which the cooling function is lower than the LTE cooling function, we use data from \citet{glover08} for a 3:1 ortho-para ratio of H$_2$. We assume H$_2$ cooling below $10^4 K$ takes place in an almost neutral medium and thus consider collisions with the following species: other H$_2$ molecules, H atoms and He atoms. Collisions with H atoms are often the only one taken into account, however, as we are interested in describing media with a high fraction of H$_2$, taking into account the collisions with H$_2$ and He is necessary to compute a cooling that would otherwise vanish in regions poor in atomic hydrogen.

The final expression for the volume cooling rate (in $\mathrm{erg} \, \mathrm{cm}^{-3} \, \mathrm{s}^{-1}$) we use in the code is:

\begin{equation}
\Lambda_{\mathrm{H}_2}= n_{\mathrm{H}_2} \frac{\Lambda_{\mathrm{H}_2, \mathrm{LTE}}}{1+\frac{\Lambda_{\mathrm{H}_2, LTE}}{n_{\mathrm{H}} \Lambda_{\mathrm{H}_2, \mathrm{H}}+n_{\mathrm{H}_2} \Lambda_{\mathrm{H}_2, \mathrm{H}_2}+n_{\mathrm{He}} \Lambda_{\mathrm{H}_2, \mathrm{He}}}}
\label{H2lamb-eq}
\end{equation}

\begin{figure}[!h]
\centering
\resizebox{\hsize}{!}{\includegraphics{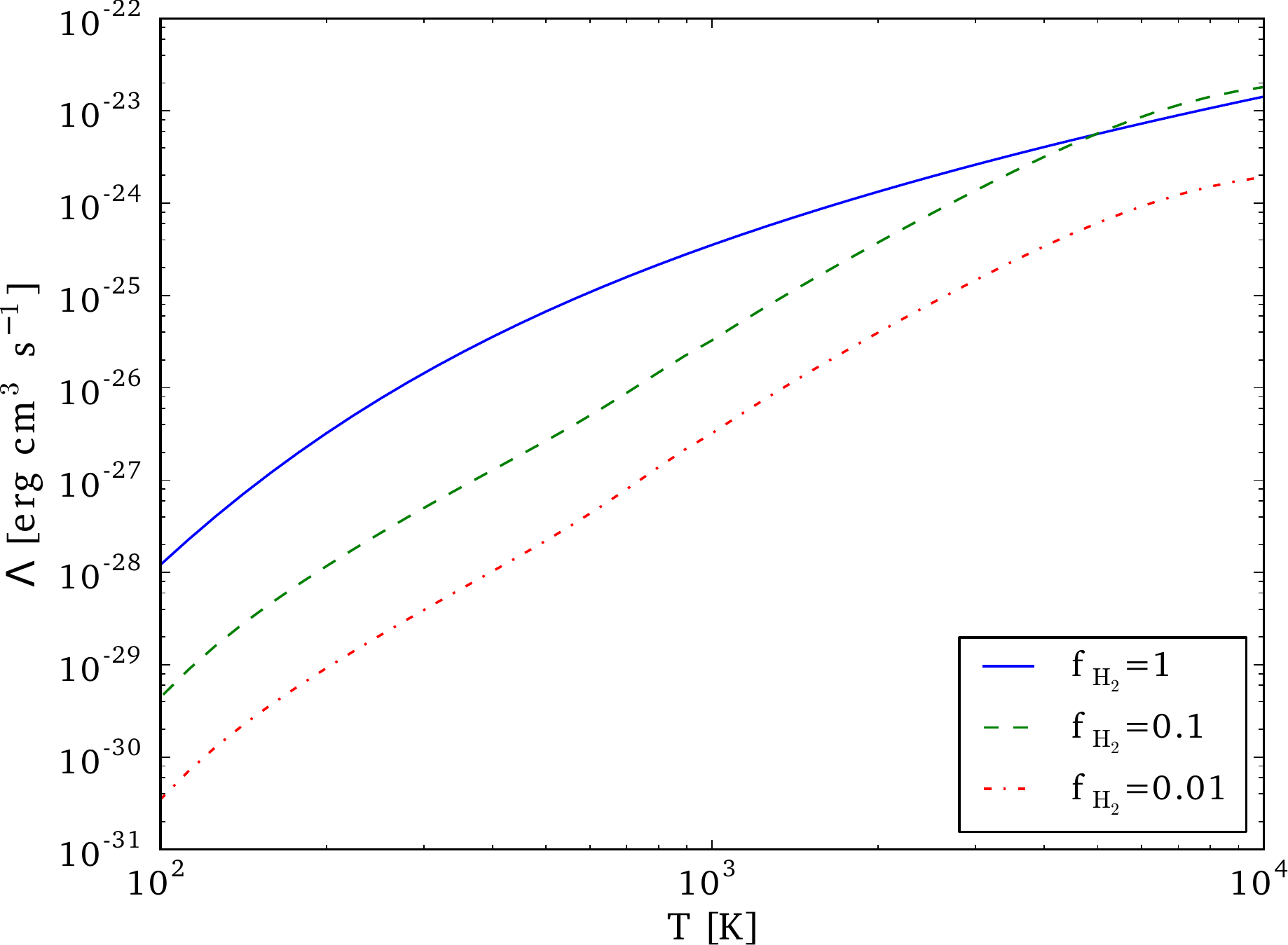}}
\caption{Volume cooling rates of H$_2$ for a hydrogen nuclei number density of 1 cm $^{-3}$ as a function of temperature for different H$_2$ to total hydrogen mass ratios $f_{\mathrm{H_2}}$. For high fractions of H$_2$ and at densities for which the cooling rate depends on collisions, the cooling can become slightly more important just below 10$^4$ K as the mass fraction is reduced, because the contribution of collisions with atomic hydrogen dominates in this range of temperature.}
\label{H2cool-fig}
\end{figure}

$\Lambda_{\mathrm{H}_2, \mathrm{LTE}}$ is the LTE cooling rate per H$_2$ molecule, in $\mathrm{erg} \, \mathrm{s}^{-1}$, and $\Lambda_{\mathrm{H}_2, \mathrm{H}}$, $\Lambda_{\mathrm{H}_2, \mathrm{H}_2}$ and $\Lambda_{\mathrm{H}_2, \mathrm{He}}$ are the cooling functions in $\mathrm{erg} \, \mathrm{cm}^{3}\, \mathrm{s}^{-1}$ due to collisions with respectively H, H$_2$ and He, for low densities. This formula, similar to what \citet{holl79} used, interpolates between the low density and high density regimes: it makes the influence of collisional processes increase with respect to the LTE cooling as the density decreases.  We plot in Figure~\ref{H2cool-fig} this cooling rate for a hydrogen nuclei number density of 1 cm$^{-3}$, and different mass ratios of molecular hydrogen over the total hydrogen component. 

We do not consider cooling due to collisions of H$_2$ molecules with metals as collisional coefficients are not well determined.

We neglect the cooling due to H$_2^+$, which abundance can be equal to the H$_2$ abundance around shocks, but then drops by order of magnitude below the H$_2$ abundance. H$_2^+$ cooling might play a role in strong starburst galaxies, or in primordial galaxies at high redshift \citep[e.g.][]{ricotti01,ahn07,petkova12}

\begin{figure*}
\centering
\resizebox{\hsize}{!}{\includegraphics{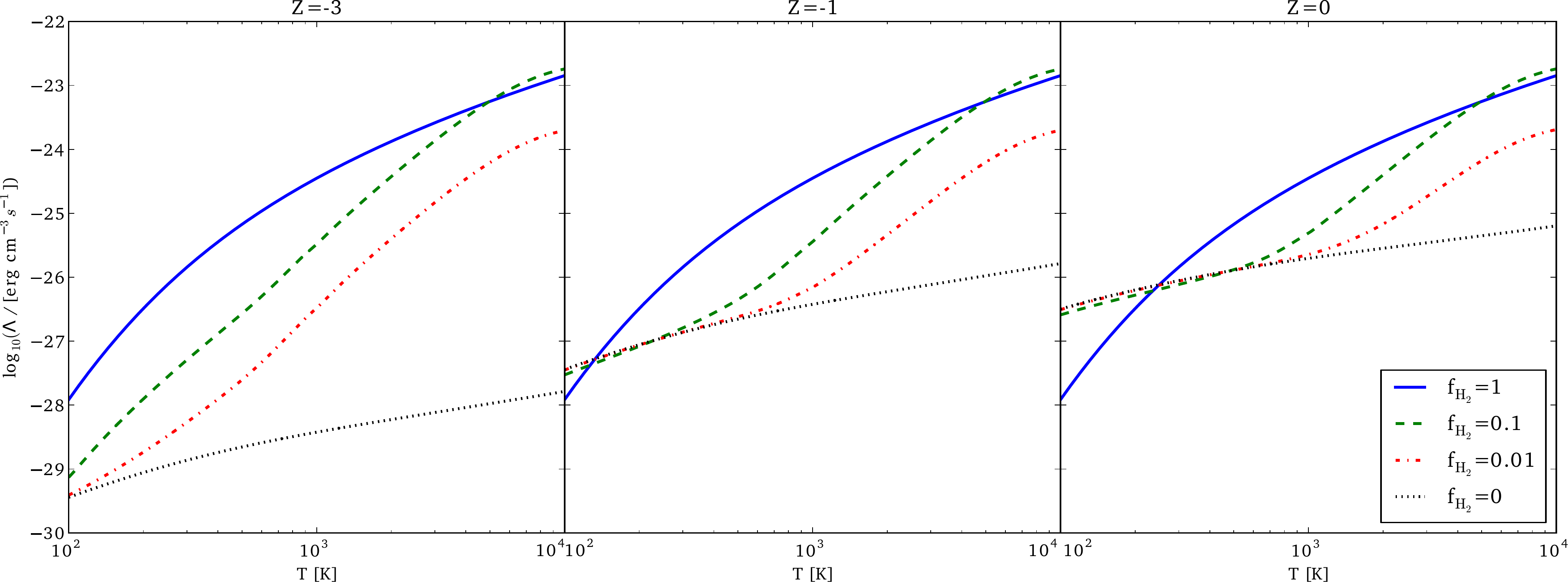}}
\caption{Volume cooling rates due to H$_2$ and metals for a hydrogen nuclei number density of 1 cm $^{-3}$ and different metallicities $Z=[\mathrm{Fe/H}]$ ($Z=0$ is solar) and H$_2$ mass fractions.}
\label{H2metcool-fig}
\end{figure*}

We show the volume cooling rate due to H$_2$ and metals for a fixed hydrogen nuclei number density $n_{\rm H_{nucl}}=1$~cm$^{-3}$ and different metallicities and H$_2$ mass fractions on Figure~\ref{H2metcool-fig}. What is plotted is $\Lambda=\Lambda_{\rm H_2}+n_{\rm H}^ 2 \Lambda_{\rm N, met}$, where $\Lambda_{\rm H_2}$ is the volume cooling rate of equation~\ref{H2lamb-eq}, and $\Lambda_{\rm N, met}$ is the volume cooling rate of Figure~\ref{cool-fig} divided by (1 cm$^{-3})^2$ (the scaling being such for these low densities of hydrogen atoms). $n_{\rm H}$ is the number density of hydrogen atoms (so that $n_{\rm H_{nucl}}= n_{\rm H}+2 n_{\rm H_{2}}$). The mass fraction of H$_2$ is $f_{\rm H_{2}}=2 \dfrac{n_{\rm H_{2}}}{n_{\rm H_{nucl}}}$. We see that H$_2$ can bring a significant contribution to the cooling even for solar metallicity gas.

\subsubsection{HD cooling}

We also consider HD cooling. Despite its lower abundance than molecular hydrogen, the molecule HD can help cool the gas because of its permanent electric dipole. The dipole transitions probabilities are orders of magnitude greater than the quadrupole ones similar to the H$_2$ quadrupole transitions, which makes the cooling per molecule significantly greater. We used quantum data to compute the LTE cooling function in a similar way than for H$_2$. For low densities, we took the function of \citet{lipov05}. This cooling function takes into account the collisions of HD with H atoms and only dipolar transitions, which is justified by their much bigger contribution to the cooling than quadrupole ones. We followed \citet{glover08} by assuming the cooling function is simply proportional to the H atoms density for low densities, so that the cooling function per molecule is, for low densities, $n_{\mathrm{H}} \Lambda_{\mathrm{HD}, \mathrm{H}} (n_\mathrm{H}=1 \, \mathrm{cm^{-3}})$

We write the volume cooling rate, as for H$_2$, as:
\begin{equation}
\Lambda_{\mathrm{HD}}= n_{\mathrm{HD}} \frac{\Lambda_{\mathrm{HD}, \mathrm{LTE}}}{1+\frac{\Lambda_{\mathrm{HD}, \mathrm{LTE}}}{n_{\mathrm{H}} \Lambda_{\mathrm{HD}, \mathrm{H}} (n_\mathrm{H}=1 \, \mathrm{cm^{-3}})}}
\end{equation}

$\Lambda_{\mathrm{HD}, \mathrm{LTE}}$ is the LTE cooling rate per HD molecule, in $\mathrm{erg} \, \mathrm{s}^{-1}$, and $\Lambda_{\mathrm{HD}, \mathrm{H}}$is the cooling function in $\mathrm{erg} \, \mathrm{cm}^{3}\, \mathrm{s}^{-1}$ due to collisions with H atoms.

In the simulations, we assume that $n_\mathrm{HD}=10^{-5} n_\mathrm{H_2}$ as the fractionation of HD with respect to H$_2$ occurs only when the gas temperature becomes as low as 100~K. The abundance can then reach 200 times the normal abundance. HD can help the cooling of the gas down to 30~K in a primordial gas without metals \citep[e.g.][]{yoshida06}. However, these temperatures are below our minimum considered. With the abundance $n_\mathrm{HD}=10^{-5} n_\mathrm{H_2}$, the influence of HD in cooling the gas is little in the simulations of this paper.


\subsubsection{Cooling algorithm}

The interstellar gas component is modelled as an ideal gas with an adiabatic index $\gamma=\dfrac{5}{3}$. Gadget-2 integrates the evolution of the specific entropy $A=\dfrac{P}{\rho^{\gamma}}$ rather than the internal energy, for computational cost and accuracy reasons developed in \citet{springel02}. 
The evolution of the specific entropy of a particle $i$ is governed by:
\begin{equation}
\frac{\dd A_i}{\dd t} = - \dfrac{\gamma - 1}{\rho_i^{\gamma}} \Lambda (\rho_i, T_i) + \dfrac{1}{2} \sum_{j=1}^{N}m_{j} \Pi_{ij}\mathbf{v}_{ij}.\mathbf{\nabla}_i \overline{W}_{ij} 
\end{equation}
where $\Lambda$ is the volume cooling rate in $\mathrm{erg} ~\mathrm{cm}^{3} \, \mathrm{s}^{-1}$ (the sum of the contributions described in above), $\Pi_{ij}$ is a factor accounting for the artificial viscosity and $\overline{W}_{ij}$ is the symmetrised SPH kernel.

As the cooling time can be lower than the time-steps equal to the minimum of $\sqrt{\frac{2 \eta \epsilon}{|\mathbf{a}|}}$ (where $\eta$ is an accuracy parameter we keep at $0.025$ as in the default Gadget-2 parametrisation, $\epsilon$ is the gravitational softening length, and $a$ is the acceleration),  and the Courant time limitation for gas particles, we use an implicit cooling scheme to stabilise the resolution. We solve iteratively the implicit equation: 
\begin{multline}
A_i(t+\dd t) = A_i(t) + \left( - \dfrac{\gamma - 1}{\rho_i(t)^{\gamma}} \Lambda (\rho_i(t), T_i(t+\dd t)) \right. \\
 + \left. \dfrac{1}{2} \sum_{j=1}^{N}m_{j} \Pi_{ij}\mathbf{v}_{ij}.\mathbf{\nabla}_i \overline{W}_{ij} \right) \dd t  
\end{multline}
where $\dd t$ is the considered time-step. This implicit scheme is however not critical for our types of simulations with little shocks and our absence of atomic/molecular heating terms.

The relation between temperature and specific entropy is:
\begin{equation}
T=\frac{\mu m_p \rho^{\gamma -1}}{k}A 
\label{T-eq}
\end{equation}
where $\mu=\Sigma_i x_i \dfrac{m_i}{m_p}$ with $x_i$ the number fraction of the species i is the mean molecular weight. For a gas composed of atomic and molecular hydrogen and helium, neglecting the metals contribution:

\begin{equation}
\mu=\frac{4}{1+\left(3-4 \dfrac{n_{\mathrm{H}_2}}{n_{\mathrm{H}_{\mathrm{nucl}}}}+4 \frac{n_{\mathrm{e^-}}}{n_{\mathrm{H}_{\mathrm{nucl}}}} \right) X  } 
\end{equation}

$n_{\mathrm{H}_{nucl}}$ is the number density of hydrogen nuclei and X is the hydrogen nuclei mass fraction that we set to 0.76. We do not compute the ionisation fraction $\frac{n_{\mathrm{e^-}}}{n_{\mathrm{H}_{\mathrm{nucl}}}}$ in the simulations. We assume the gas is quasi neutral below $10^4$ K. Above that temperature, we assume there is no molecular hydrogen and the gas is fully ionised (H$^+$ and He$^{2+}$). The mean molecular weight being higher for a neutral gas than for an ionised one, there is a specific entropy range for which the temperature obtained (from equation~\ref{T-eq}) by assuming a neutral gas is above 10$^4$~K while the temperature obtained by assuming an ionised gas is below 10$^4$K. We set to $10^4$ K the temperature of the gas in this specific entropy range.

\subsection{Resolution}

\begin{table}
\caption{Resolution}
\begin{flushleft}
\begin{tabular}{cccc}
\hline
$\epsilon$ & m$_{\mathrm{DM}}$ & m$_{\star}$ & m$_{g}$  \\
$[\mathrm{pc}]$ & [M$_\odot$]  &[M$_\odot$]  &[M$_\odot$] \\
\hline
100 & 3.7 $10^{5}$ & 1.4 10$^{5}$ & 2.5 10$^{4}$    \\
\end{tabular}
\end{flushleft}
\caption{Resolution: gravitational softening and particle masses.}
\label{t-res}
\end{table}

Distinguishing physical effects from numerical ones can be difficult if the resolution of a simulation is not well adapted. The number of particles setting the mass resolution, the gravitational softening length and the hydrodynamics smoothing length need to be chosen so that they do not introduce unwanted artefacts while allowing for the simulation to run, and at a reasonable pace. 

We want to describe the gas down to low temperatures, but if the temperature is too low, the Jeans length and Jeans mass of the gas will be more poorly resolved, which can lead to numerical artefacts \citep{bate97}. \citet{schaye08} use an effective equation of state at high densities that makes the Jeans mass independent of density, while \citet{hopkins11} take a density-dependent pressure floor to ensure the Jeans length is resolved, and \citet{bour10} use an equation of state that mimics the effect of cooling at all temperatures and a temperature floor at high densities. We take a temperature floor of 100~K for all densities. 
We note that since we simulate disc galaxies, the gas is stabilised by the rotation of the disc. 

The number of particles should ideally be the largest possible to have a good mass resolution, which allows the Jeans mass to be well resolved. We run simulations with 1\,200\,000 initial particles: a third are gas particles, a third are stellar particles and a third are dark matter particles. Table~\ref{t-res} shows the particle masses we have in the particular case of our galaxy model that will be detailed in \ref{initcond}. New stellar particles have a mass $m_{\mathrm{new~*}}=\frac{m_g}{N_{g\rightarrow\star}}$ depending on the parameter $N_{g\rightarrow\star}$ described in \ref{starform}, the number of stellar particles produced out of one gas particle.

The gravitational softening length $\epsilon$ depends on this number of particles: a too small value for a given number of particles introduces unphysical two-body relaxation in media that are collisionless (stellar components of galaxies and dark matter haloes), while a too large one decreases the spatial resolution by ``blurring'' density features and does not allow the Jeans length to be gravitationally resolved. Gravitational softening lengths $\epsilon$ are usually taken as scaling with the mean inter-particle distance, therefore as the number of particles to the power $\frac{1}{3}$ for a 3-dimensional simulation. We take a softening length $\epsilon=100$~pc for all particle types. This particular value is derived from the GalMer simulations \citep[eg][]{dimatteo07} which took ten times less particles for a softening of 280~pc. The GalMer simulations were isothermal, at $10^4$ K, and since here the temperature reaches down to 100~K, we take a softening length that is a little inferior to the cubic root, while still allowing for an efficient computation on a few tens of computing cores.

The value of the gas smoothing length $h$ should be small to allow for a good density resolution. Gadget-2 uses variable smoothing lengths: the densities and smoothing lengths are computed together, so that the mass contained in a sphere of radius $h$ is a constant. We take a minimum $h$ equal to a tenth of the gravitational softening. This minimum can have subtle consequences on the structure of the gas.
In the presence of strong cooling and with no sufficient star formation or efficient feedback, a number of particles can 
be ``stuck'' at this minimum, making the corresponding regions increasingly denser with no possibility of getting more diffuse. 
This can be computationally very demanding and lead to large domain work-load imbalance.

\subsection{H$_2$ fraction} 

Our goal is not to determine the molecular abundance through a detailed chemical scheme, which would be too sophisticated for our present approximated treatment (which does not take into account radiative transfer). \citet{kmtI,kmtII,kmtIII} have derived an analytic expression of the mass fraction of molecular hydrogen $f_{\mathrm{H}_2}$ in an idealised spherical cloud submitted to a uniform and isotropic radiation field. Including in their study radiative transfer and the formation and destruction of H$_2$ through dust, assuming a steady state, they obtain:
\begin{equation}
f_{\mathrm{H}_2}\simeq 1-\frac{3}{4}\frac{s}{1+0.25s}
\label{fh2-eq}
\end{equation}
with:
\begin{equation}
s=\frac{\ln(1+0.6 \chi + 0.01 \chi^2)}{0.6 \tau_C}
\label{s-eq}
\end{equation}

$\tau_C$ is the dust optical length and $\chi$ is a scaled UV flux, divided by the hydrogen nuclei number density. $f_{\mathrm{H}_2}$ is set to zero when $s>2$. 

The dust optical length is $\tau_C=\dfrac{\Sigma \sigma_d}{\mu_\mathrm{H}}$. The dust cross section to the Lyman-Werner radiation per H nucleus $\sigma_d$ is set, using the reference Milky Way value, to $\sigma_d=Z' 10^{-21} \, \mathrm{cm}^{2}$ \citep[e.g.][]{krum11}, with $Z'$ the metallicity we normalise by the solar metallicity. $\mu_\mathrm{H}$ is the mean mass per H nucleus. $\Sigma$ is a column density $\Sigma=\rho L$ obtained from a local scale height: $L=\dfrac{\rho}{|\nabla \rho|}$. We compute $(\nabla \rho)_i$, the gradient of the density of the particle $i$ by: $\rho_i(\nabla \rho)_i=\sum_j m_j(\rho_j-\rho_i) \nabla_i W(r_{ij}, h_i)$. The scale height takes the variation of the density into account: it increases with density but is inversely proportional to its gradient, so it is lower in the case of large gradients encountered on the outer parts of density features like clumps or spiral arms. It is thus well adapted to compute the column density used to determine the shielding from radiation. 
 
The $\chi$ is computed using the new stars radiation field $G$. We assume stars whose age is inferior to 10~Myr radiate in the Lyman-Werner band and can dissociate the H$_2$ molecules, and do not consider that older stars radiate in this band, so they don't have any effect on the fraction of H$_2$ in our simulations. We take $\chi \propto \dfrac{G}{{n_\mathrm{H}}_{\mathrm{nuclei}}}$, with a scaling factor depending on the resolution of the simulations, and tuned so that the obtained H$_2$ fractions are consistent with observations. 
We do not follow the radiative transfer of the photons from young stars. The radiation flux produced by stars decreases with the distance $r$ to a star in $\frac{1}{r^2}$, as the gravitational field (We assume an ISM escape fraction equal to unity.). We insert the computation of the flux received from stars younger than 10~Myr by one gas particle in the Gadget-2 gravitational tree functions, as it is similar to computing the gravitational force due to the young stars only and can be easily done by adding a range of variables representing this contribution only. We thus use the gravitational tree to compute a flux proportional to the mass of new stars over the squared distance.

\subsection{Star formation} 
\label{starform}

We implement star formation in a stochastic way that reproduces a Schmidt law \citep{katz92}:
\begin{equation}
\frac{\dd \rho_{*}}{\dd t}=-\frac{\dd \rho_{g}}{\dd t}= \frac{c_{\star}}{t_{\mathrm{ff}}} \rho_g
\end{equation}
where $\rho_{*}$ is the volume stellar density, $\rho_{g}$ the volume gas density and $t_{\mathrm{ff}}$ is the free fall time: $t_{\mathrm{ff}}=\sqrt{\frac{3 \pi}{32 G \rho_g}}$, and $c_{\star}$ is the star formation efficiency per free fall time.

The Schmidt law can be enforced by giving a gas particle a probability to spawn a star at each time-step $\Delta t$:
\begin{equation}
p_*=\frac{m_g}{m_*}\left(1-e^{-\frac{c_{\star}}{t_{\mathrm{ff}}}\Delta t}\right)
\end{equation}

This implementation means that a gas particle of mass $m_g$ can spawn a star particle of mass $m_*$ with this probability at each time-step, the mass of the gas particle being then reduced by the amount of $m_*$, until there is no more mass in the gas particle. The number of $N_{g\rightarrow\star}$ of star particles created by gas particle is a compromise between a good mass and time resolution of star formation and CPU cost: if several stellar particles are created from a single gas particle, star formation will be smoother, temporally better resolved but the total number of particles can increase significantly, which slows down the code. In the case where $N_{g\rightarrow\star}$ is greater than one, we note that this spawning scheme implies that gas particles will have different masses, either the initial one $m_g$, or a smaller multiple of $\frac{m_g}{N_{g\rightarrow\star}}$. The density and smoothing length are computed by Gadget-2 so that the mass contained in a sphere of radius $h_i$ is fixed, $h_i$ being the smoothing length of the particle $i$, and we were careful about modifying the algorithm to take into account the different masses, so that this mass condition is still satisfied. 

We use common selection rules for the particles that are allowed to spawn stars: they must have a density higher than a threshold ${n_\mathrm{H}}_{\mathrm{min}}$, must have a temperature lower than a maximum temperature $T_{\mathrm{max}}$ and must be in a converging flow ($\mathbf{\nabla . v}<0$). 
We set ${n_\mathrm{H}}_{\mathrm{min}}=10^{-1}cm^3$. The threshold can be increased to allow for an ISM with more density structures, but if it is too high, the mass resolution must be increased for the Jeans mass to still be resolved. 
We use $T_{\mathrm{max}}=30\,000$~K, which happens not to be selective as only a few diffuse particles can reach these temperatures in our simulations.

\subsection{Feedback} 

Our simulations include core-collapse supernovae kinetic feedback. 
We consider a supernova explosion releases the canonical value $E_{\mathrm{SN}}=10^{51} \mathrm{erg}$. A fraction $\alpha_{\mathrm{fb}}$ of this energy is given to the ISM, while the rest is radiated away. Each new stellar particle of mass $m_{\star}$ inputs an energy $E_{\mathrm{input}}= \alpha_{\mathrm{fb}} \epsilon_{\mathrm{SN}} m_{\star}$ where $\epsilon_{\mathrm{SN}}$ is the number of formed supernovae per formed stellar mass multiplied by the canonical energy $E_{\mathrm{SN}}=10^{51} \mathrm{erg}$, i.e. the supernovae energy released per formed stellar mass. We consider this mass fraction is 0.01, which is of the order of the fractions obtained from commonly used initial mass functions (such as a Salpeter IMF with slope $-$1.35 and lower and upper limits of 0.1 and 40 $\mathrm{M_{\odot}}$, with stars more massive than 8 $\mathrm{M_{\odot}}$ considered to be supernovae), so we have $\epsilon_{\mathrm{SN}}=10^{49} \mathrm{erg/ M_{\odot} }$.

Each neighbour $i$ of a new star particle $0$ receives an energy weighted by its distance to the new stellar particle: 
\begin{equation}
E_i= \frac{W(|\mathbf{r_{i0}}|,h_0)}{\sum_{\text{ngb~k}}{W(|\mathbf{r_{k0}}|,h_0)}} E_{\mathrm{input}}
\end{equation}
so that the sum of the energies given to the neighbours is $E_{\mathrm{input}}$. If the newly created stellar particle has left a remnant gas particle, no feedback energy is given to the remnant. We recompute the smoothing length and neighbours list at the position of a new 
stellar particle, considering only the neighbouring gas that has not been turned into stars (and not considering a possible gas particle remnant at the exact position of the new stellar particle). Therefore the sum of the fractions involving kernels is indeed unity, and the mass of the gas affected by each supernova explosion is a constant. 

The neighbours are given a velocity kick $\sqrt{\frac{2 E_i}{m_i}}$, directed along the line joining the stellar particle and the neighbour, and away from the new stellar particle. The feedback is only kinetic. This input kinetic energy can however be converted into internal energy through the SPH viscosity.

The number $N_{g\rightarrow\star}$ of stellar particles created out of one gas particle has an influence on the distribution of feedback energy. Feedback is more gradually input for a larger $N_{g\rightarrow\star}$. 

\subsection{Initial Conditions} 
\label{initcond}

We consider the case of a giant Sb galaxy. 

The gaseous and stellar discs follow Miyamoto-Nagai density profiles. The density of the gas disc is, in cylindrical coordinates:
\begin{multline}
\rho_g(R,z)=\frac{h_g^2 M_g}{4\pi} \\ 
\times \frac{a_g R^2+\left( a_g+3 \sqrt{z^2+h_g^2}\right) \left( a_g+\sqrt{z^2+h_g^2} \right)^2  }{ \left( R^2+ \left( a_g+ \sqrt{z^2+h_g^2}\right)^2  \right)^ {\frac{5}{2}} \left(z^2+h_g^2 \right)^{\frac{3}{2}} }
\end{multline}
and the stellar disc density is:
\begin{multline}
\rho_d(R,z)=\frac{h_d^2 M_d}{4\pi} \\ 
\times \frac{a_d R^2+\left( a_d+3 \sqrt{z^2+h_d^2}\right) \left( a_d+\sqrt{z^2+h_d^2} \right)^2  }{ \left( R^2+ \left( a_d+ \sqrt{z^2+h_d^2}\right)^2  \right)^ {\frac{5}{2}} \left(z^2+h_d^2 \right)^{\frac{3}{2}} }
\end{multline}

The Miyamoto-Nagai profiles are chosen because the associated potential is analytic and the initial velocity dispersions can thus be easily computed, but relaxation makes the profiles quickly exponential as observed.

We use the gravitation routines of Gadget-2 to obtain the forces acting on particles. We compute the circular velocity of disc particles using these gravitational accelerations, and add an analytic asymmetric drift correction to have a more realistic velocity profile. 

The radial velocity dispersion is derived from $\sigma_r(r)=\frac{3.36 G Q\Sigma(r)}{\kappa(r)}$, with Q the Toomre parameter that we set to 1 for both discs, $\Sigma$ the surface density and $\kappa$ the epicyclic frequency derived from the potential. The azimuthal velocity dispersion is obtained from $\frac{\sigma_{\theta}(r)}{\sigma_r(r)}=\frac{\kappa(r)}{2\Omega(r)}$ with $\Omega$ the angular speed, and the vertical dispersion is set by the isothermal equilibrium of the disc. 

The spherical stellar bulge and dark matter halo have Plummer density profiles. The bulge density is thus given by:
\begin{equation}
\rho_b(r)=\frac{3M_b}{4\pi r_b}\left( 1+\frac{r^2}{{r_b}^2}\right)^{-\frac{5}{2}} 
\end{equation}
and the halo density is:
\begin{equation}
\rho_h(r)=\frac{3M_h}{4\pi r_h}\left( 1+\frac{r^2}{{r_h}^2}\right)^{-\frac{5}{2}} 
\end{equation}

All the masses and characteristic lengths are specified in Table~\ref{t-sim}.

The velocity dispersion of the spherical components is chosen to be isotropic and is derived from the second moment of the Jeans equation. The radial dispersion is thus:
\begin{equation}
\sigma^2_r(r)=\frac{1}{\rho (r)}\int_r^\infty \rho (u) \frac{\partial \phi}{\partial u} \dd u
\end{equation}

\begin{figure}[!h]
\centering
\resizebox{\hsize}{!}{\includegraphics{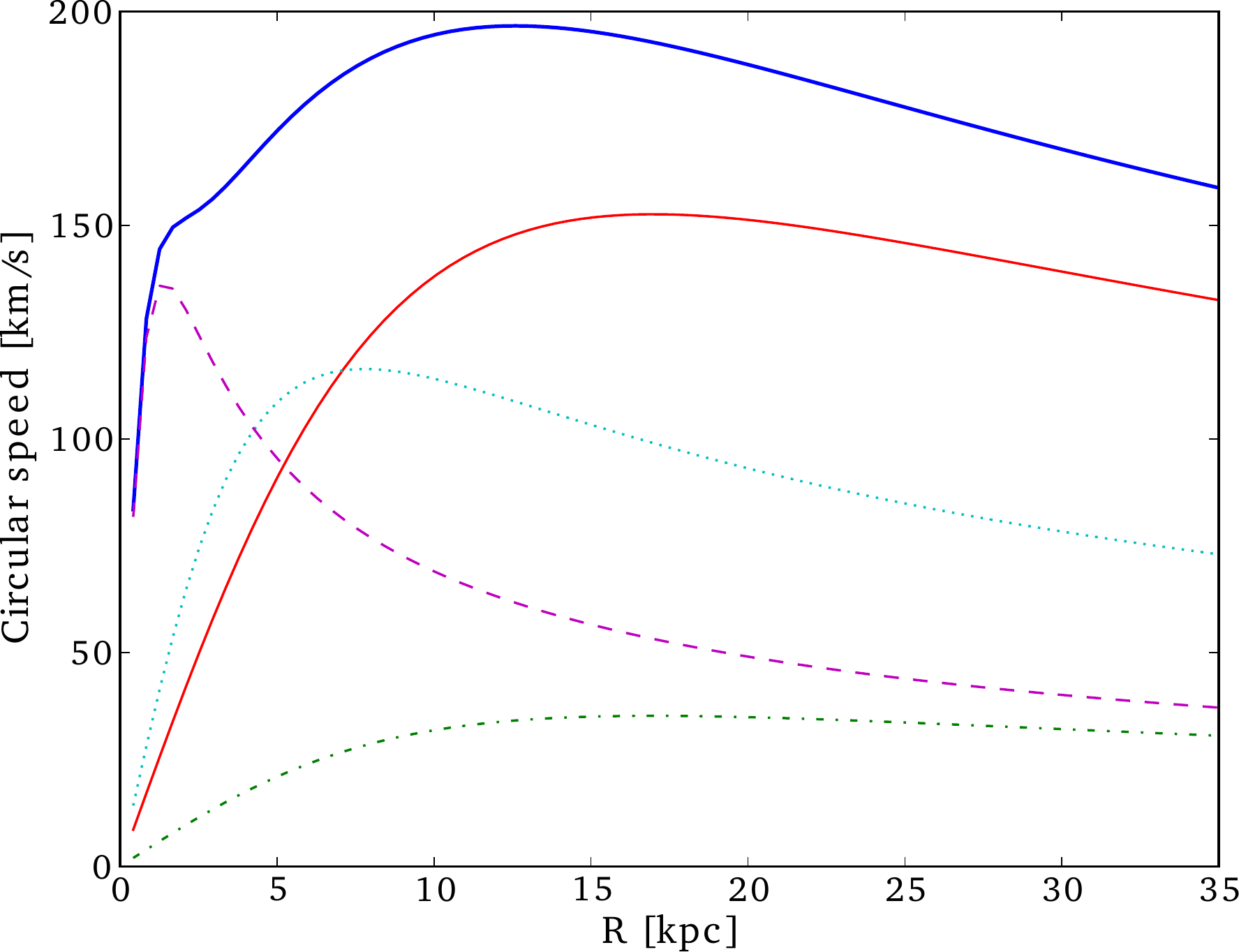}}
\caption{Initial rotation curve. Top solid (blue) thick line: total rotation curve. Dashed (purple) line: bulge. 
Dotted (green) line: stellar disc. Dash-dot (green) line: gas. Solid (red) line: dark matter halo}
\label{vc-fig}
\end{figure}

The velocity curve for this analytic model, with the contributions of the different components, is shown in Figure~\ref{vc-fig}. The rotation is due mainly to  the stellar components near the center of the galaxy, and to dark matter at large radii.

\begin{figure}[!h]
\centering
\resizebox{\hsize}{!}{\includegraphics{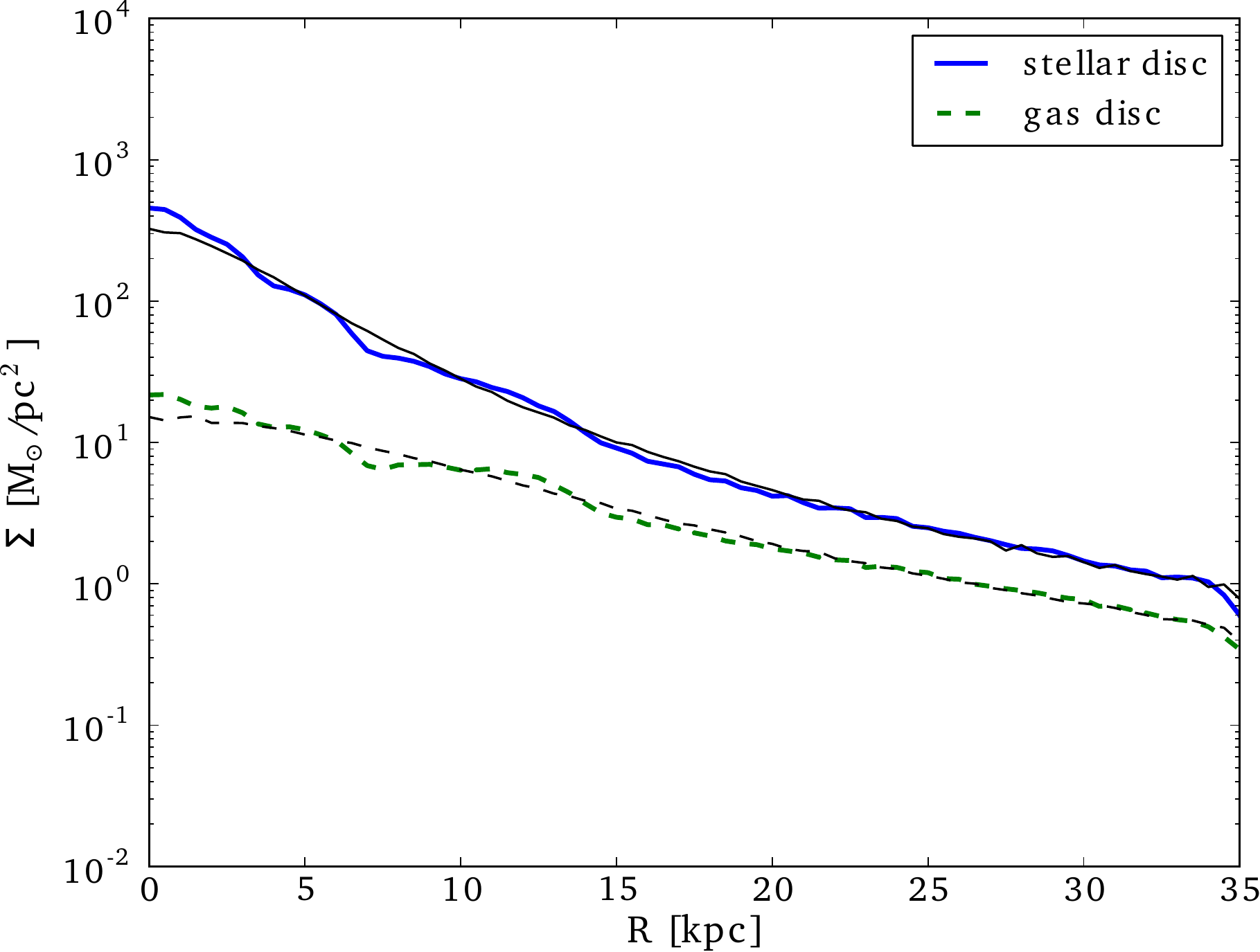}}
\caption{Initial surface densities after 300~Myr of initial conditions preparation. Dotted green line: gas disc. Dashed blue line: stellar disc. The black lines (dotted for gas and solid for stars) represent the analytic profiles taken at the beginning of the initial conditions preparation.}
\label{surfrho-fig}
\end{figure}

\begin{figure}[!h]
\centering
\includegraphics[width=7.5cm]{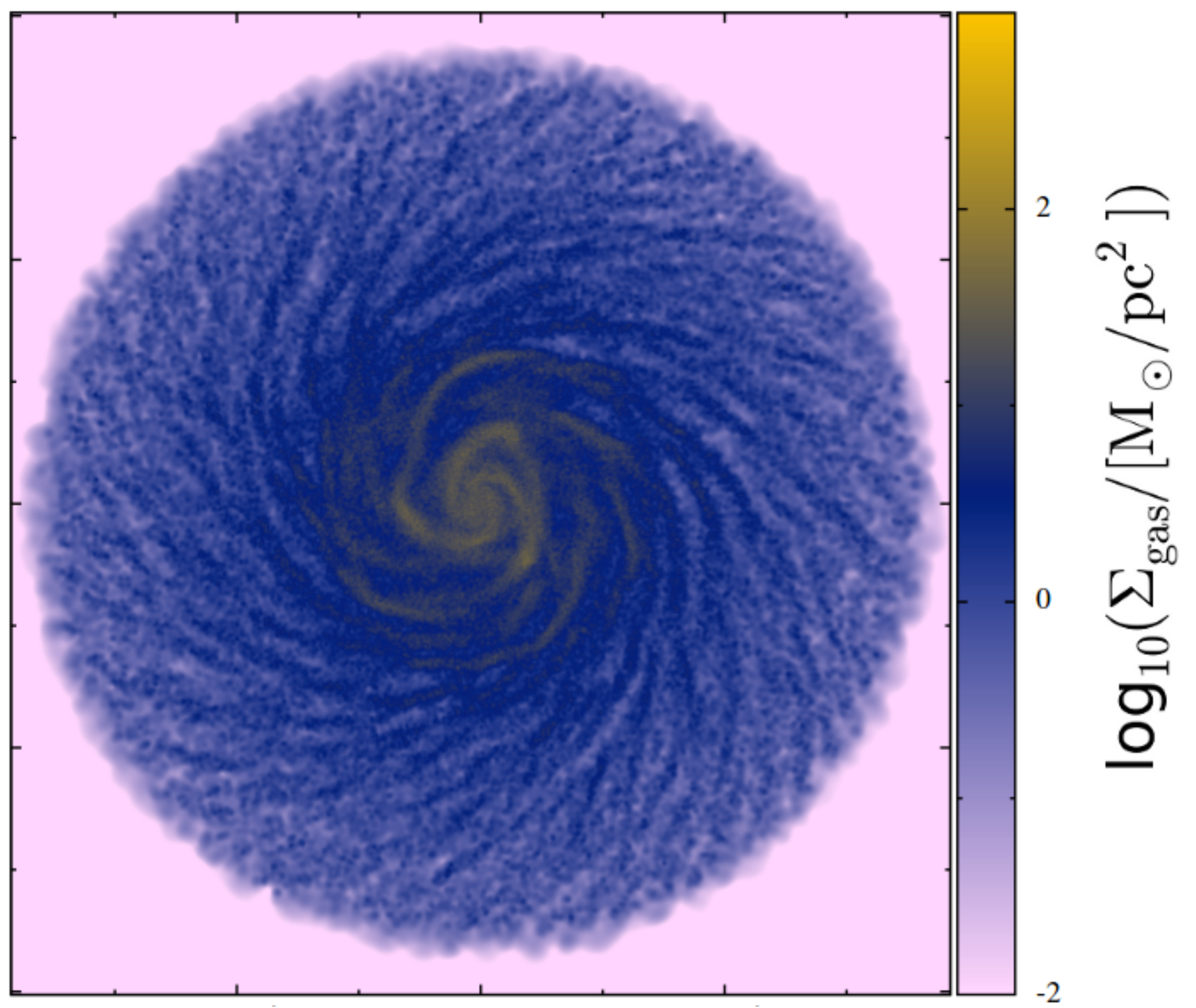}
\caption{Initial density map of the gas disc after 300~Myr of initial conditions preparation. The box size is [40~kpc x 40~kpc]}
\label{initgas-fig}
\end{figure}

We further prepare the initial conditions by letting the galaxy evolve for 300~Myr with only gravitational forces included for any type of particles. This allows us to start the simulations with dynamically relaxed discs that will especially exhibit no annuli instabilities. The initial surface densities of gas and stars are shown in Figure~\ref{surfrho-fig}, and a snapshot of the gas is shown on Figure~\ref{initgas-fig}.

\begin{table}
\caption{Sb galaxy parameters}
\begin{flushleft}
\begin{tabular}{lccccc}
\hline
Sb & M$_{h}$ & M$_{d}$ & M$_{b}$ & M$_{g}$ &   \\
\hline
in  M$_\odot$  & 1.7 $10^{11}$ & 4.5 10$^{10}$ & 1.1 10$^{10}$ & 0.9 10$^{10}$ &   \\
\end{tabular}
\begin{tabular}{lcccccc}
\hline
       &  $r_h$ & $a_d$ & $h_d$ & $r_b$ & $a_g$ & $h_g$ \\
\hline
in kpc & 12  & 5  & 0.5 & 1 & 11.8 & 0.2

\end{tabular}
\end{flushleft}

\label{t-sim}
\end{table}

\section{Simulations and results} 
\label{resu}

We assume there is a metallicity gradient in the gas. We take a central metallicity $Z=[\mathrm{Fe/H}]$ such that $Z (R=0) = Z_{\odot} + 0.5$ and assume the metallicity decreases of 1~dex per 10~kpc:
\begin{equation}
Z(R) = Z(R=0) - \dfrac{R[\mathrm{kpc}]}{10}
\end{equation}
where $R$ is the cylindrical radius. The influence of the cooling by metals will thus decrease with the distance to the centre of the galaxy. For simulations with no molecular hydrogen, we expect the outer regions to have fewer density features than the central ones and to be warmer. If however we add some molecular hydrogen, depending on its fraction and on the details of star formation and feedback, the gas may be able to form clumps also in the outer regions and form more stars there. 

The gas temperature is initially 100~K. The temperature floor we apply is more exactly a specific energy floor, corresponding to a temperature of 100~K for purely atomic gas, and going up to 186~K for gas with hydrogen present only in the molecular form.

For all the simulations of the paper, unless otherwise specified, the star formation efficiency per free-fall time is set to $c_\star=0.1$, and the number density threshold for star formation is ${n_\mathrm{H}}_{\mathrm{min}}=10^{-1}$ cm$^3$. We show some consequences of a variation of these parameters in section \ref{H2-sec}
. The simulations presented here all have a number of stars spawned by gas particle $N_{g\rightarrow\star}~=~4$.

\subsection{Simulations without H$_2$}

\begin{figure*}
\centering
\includegraphics[width=17.5cm]{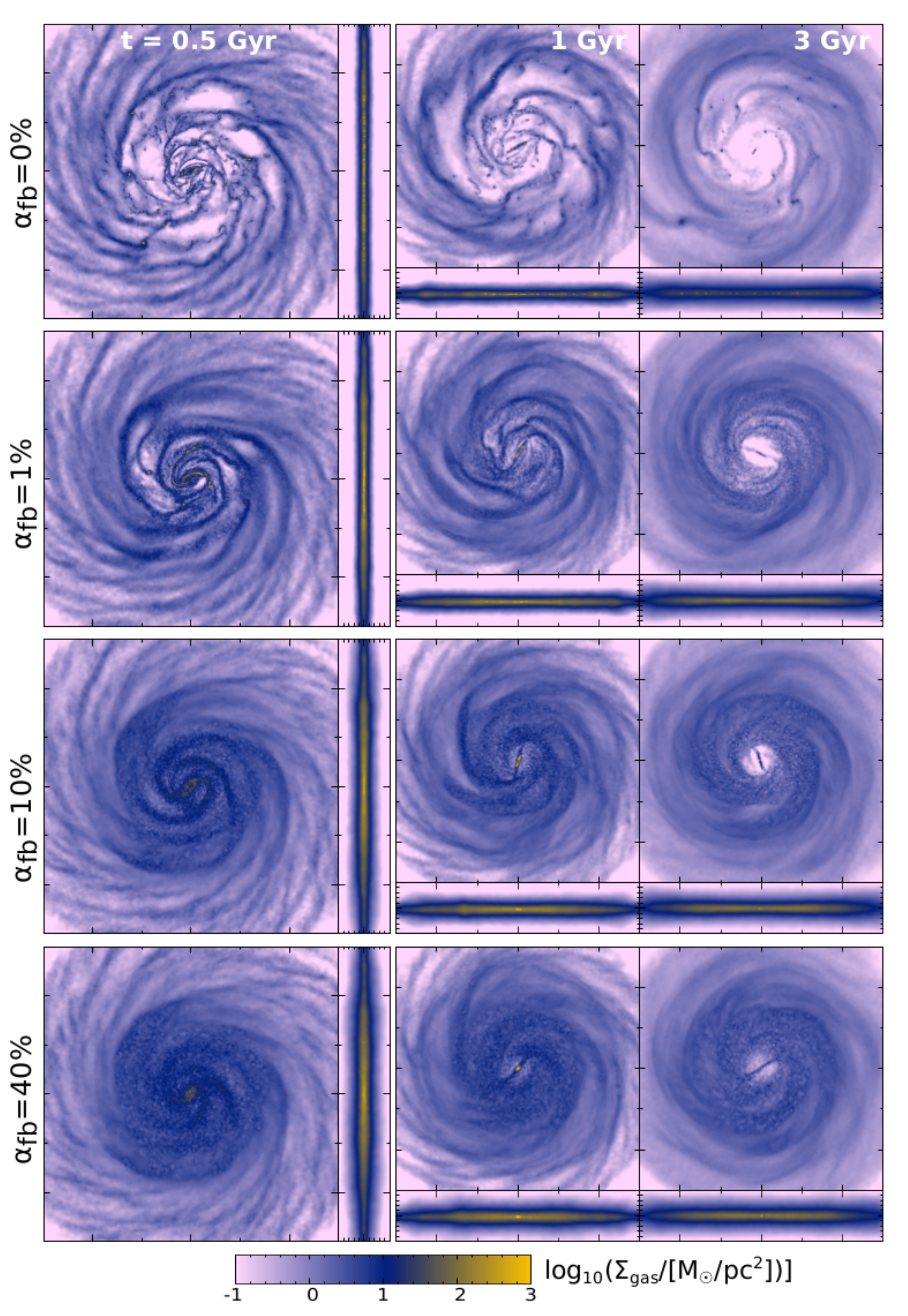}
\caption{Projections of the gas density after 0.5 Gyr, 1 Gyr and 3 Gyr of evolution. Box sizes are [60~kpc x 60~kpc] for face-on views and [60~kpc x 10~kpc] for edge-on views. Each row corresponds to a feedback efficiency indicated on the left. The column integrated density scale is the same for all plots. Done with the visualisation software SPLASH \citep{price07}.}
\label{aspectnoH2-fig}
\end{figure*}

We first perform simulations with no molecular hydrogen, with varying feedback efficiencies $\alpha_{\mathrm{fb}}$. We have four different feedback efficiencies: either no feedback ($\alpha_{\mathrm{fb}}=0$), $\alpha_{\mathrm{fb}}= 1 \%$, $\alpha_{\mathrm{fb}}= 10 \%$ or $\alpha_{\mathrm{fb}}= 40 \%$. 

Density maps of the gas discs for these runs are shown in Figure~\ref{aspectnoH2-fig} at three simulation times: 0.5~Gyr, 1~Gyr and 3~Gyr.  The gas can reach smaller values of local volume density when feedback is included, as star formation in dense regions inputs some energy in the ISM and prevents it from getting denser, slowing down the star formation. For the run without feedback, the central part of the galaxy exhibits a strong density contrast on the snapshots at 0.5~Gyr and 1~Gyr, many small clumps and thin spiralling filaments can be seen. If feedback is switched on, we can see only a few clumps for a feedback efficiency $\alpha_{\mathrm{fb}}=1 \%$ and smoother spiral features, and no clumps for $\alpha_{\mathrm{fb}}=10 \%$ and $\alpha_{\mathrm{fb}}=40 \%$. A central thin bar is formed in all the cases. After 3~Gyr, the bar is still clearly visible for the higher feedback efficiencies, but is less obvious otherwise, because the central parts have been depleted from gas by star formation.

\begin{figure}
\centering
\includegraphics[width=7.5cm]{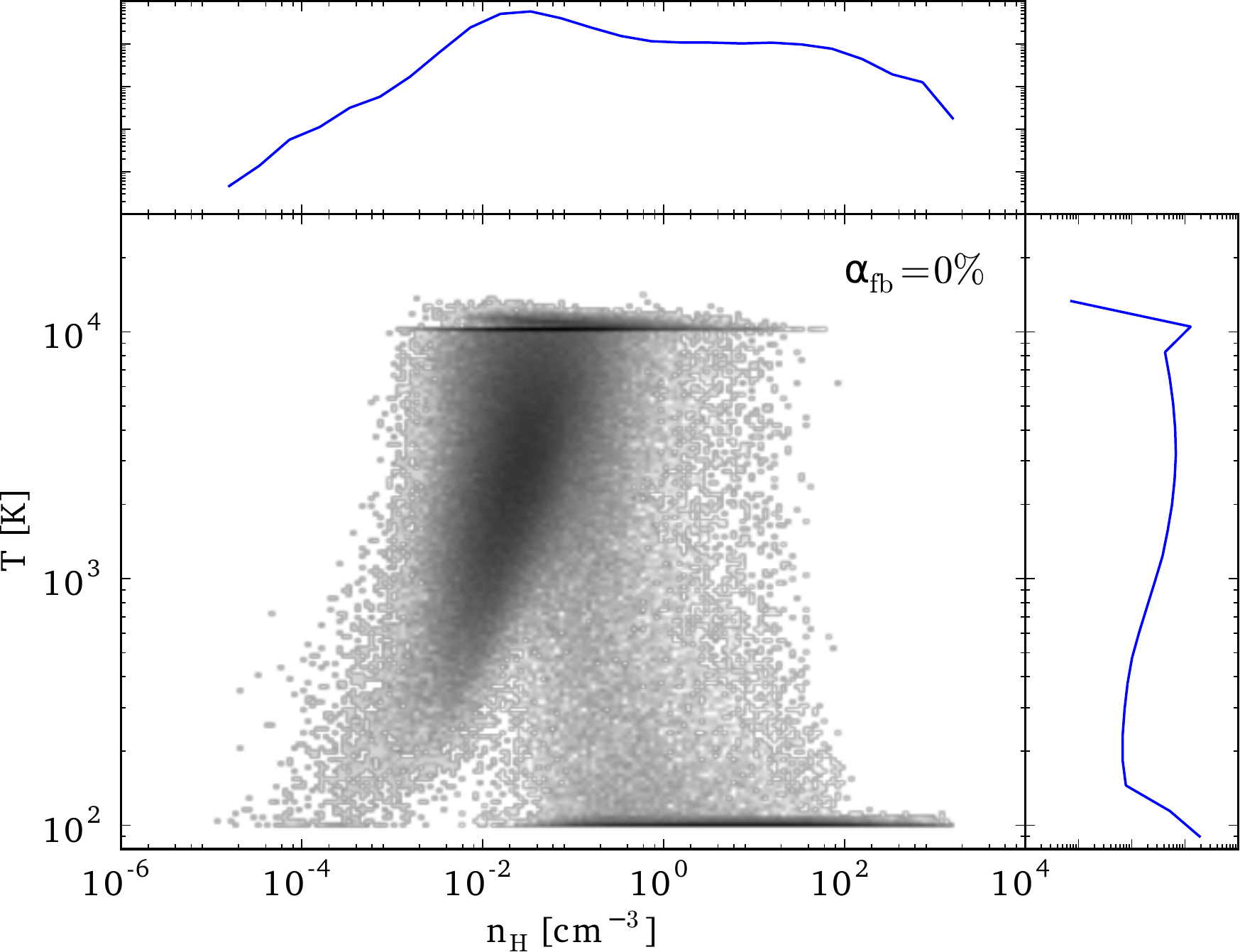}
\includegraphics[width=7.5cm]{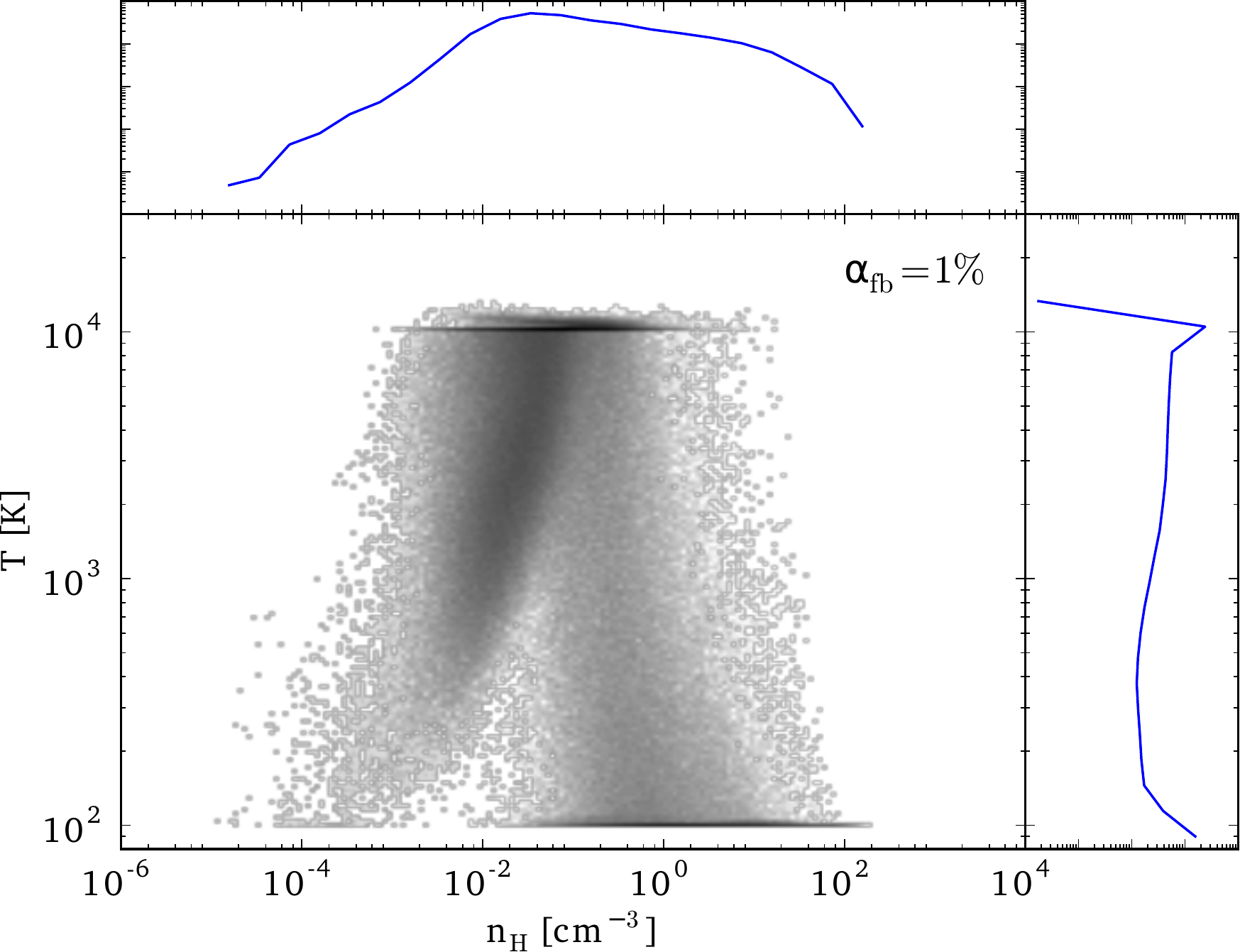}
\includegraphics[width=7.5cm]{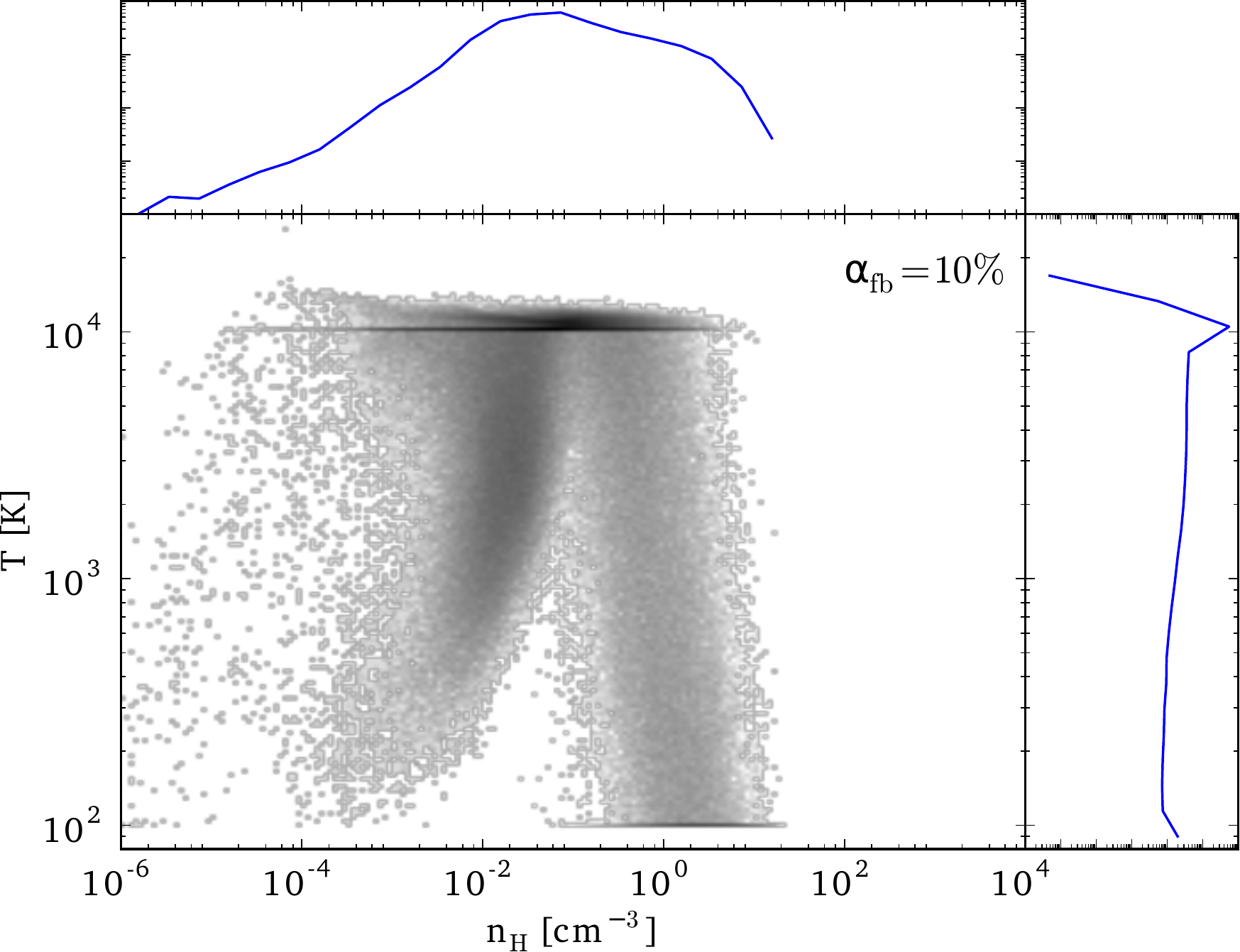}
\includegraphics[width=7.5cm]{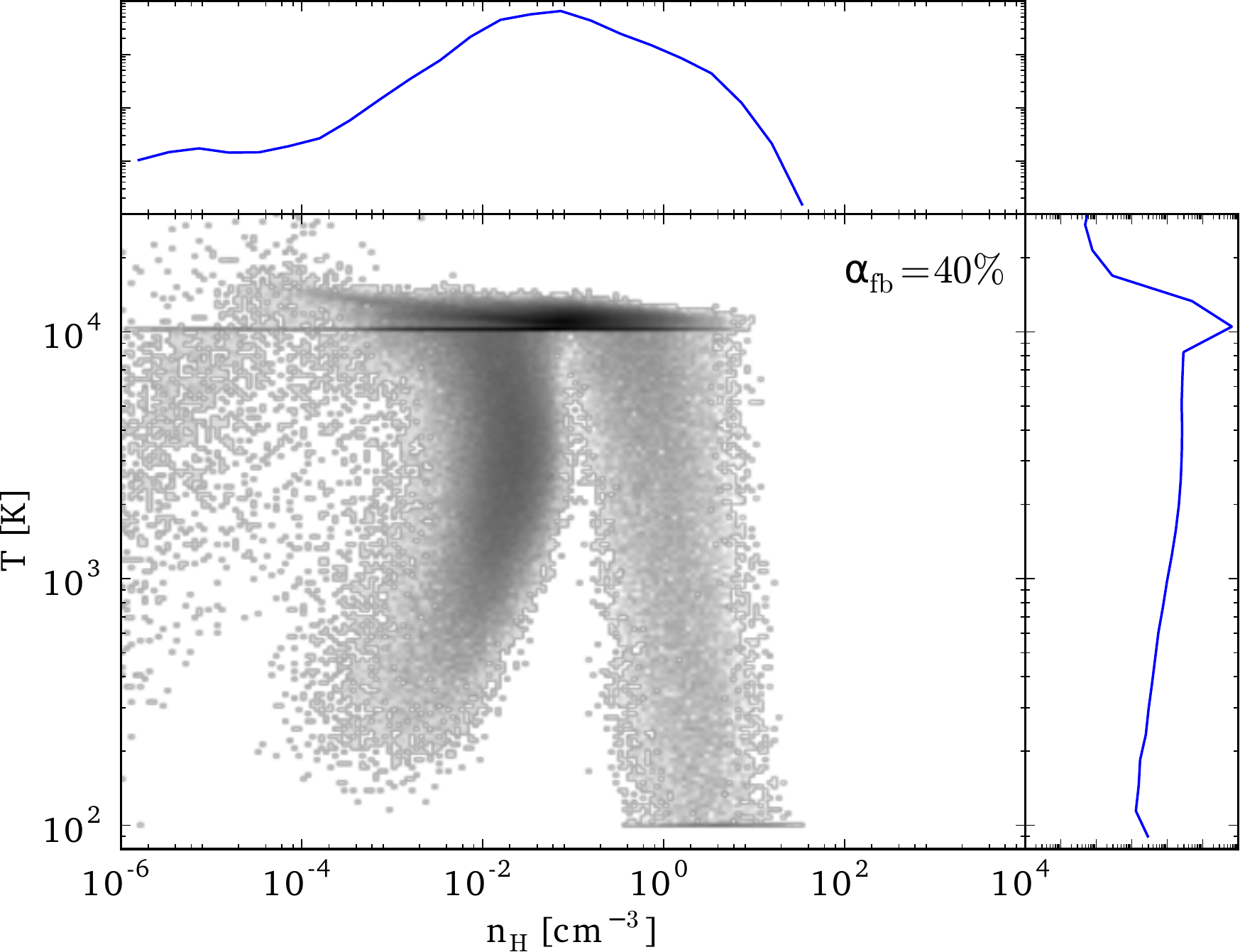}
\caption{Temperature-hydrogen number density histogram of the gas after 0.5 Gyr of evolution. The 2D histograms and marginal 1D histograms are all mass weighted and normalised. From top to bottom: runs with an increasing feedback efficiency.}
\label{hrTnoH2-fig}
\end{figure}

Figure~\ref{hrTnoH2-fig} shows the specific energy-number density histograms after 0.5~Gyr of evolution, a time at which gaseous discs are actively forming stars in all the simulations. On the top and right of each plot, the marginal probability density functions (PDFs) of respectively hydrogen nuclei number density and specific energy are shown. The specific energy PDFs show the fraction of gas at T~$\simeq 10^4$~K (corresponding to the higher specific energy concentration) increases with feedback, while the cold dense gas fraction decreases. On the density PDFs we can see that gas reaches smaller maximum densities for higher feedback efficiencies and there is an increasingly high fraction
 of diffuse gas. In the low feedback runs, a significant fraction of the dense gas has a temperature close to the minimum: this is the dense gas of the central parts, subject to metal-line cooling. The fraction decreases for higher feedback efficiencies, because of the dissipation of energy by feedback, making the densest gas of the simulations warmer. If the gas is heated by pressure forces, viscous shocks or feedback, its temperature does not reach 
much beyond 10~$^4$K because of the stronger H, He, and metals cooling it undergoes above. The diagonal branches observed on the left of each plot account for gas that, away from the centre of the disc, is subject to very little cooling because of the metallicity gradient, and thus cools down adiabatically. 

\begin{figure}[!h]
\centering
\resizebox{\hsize}{!}{\includegraphics{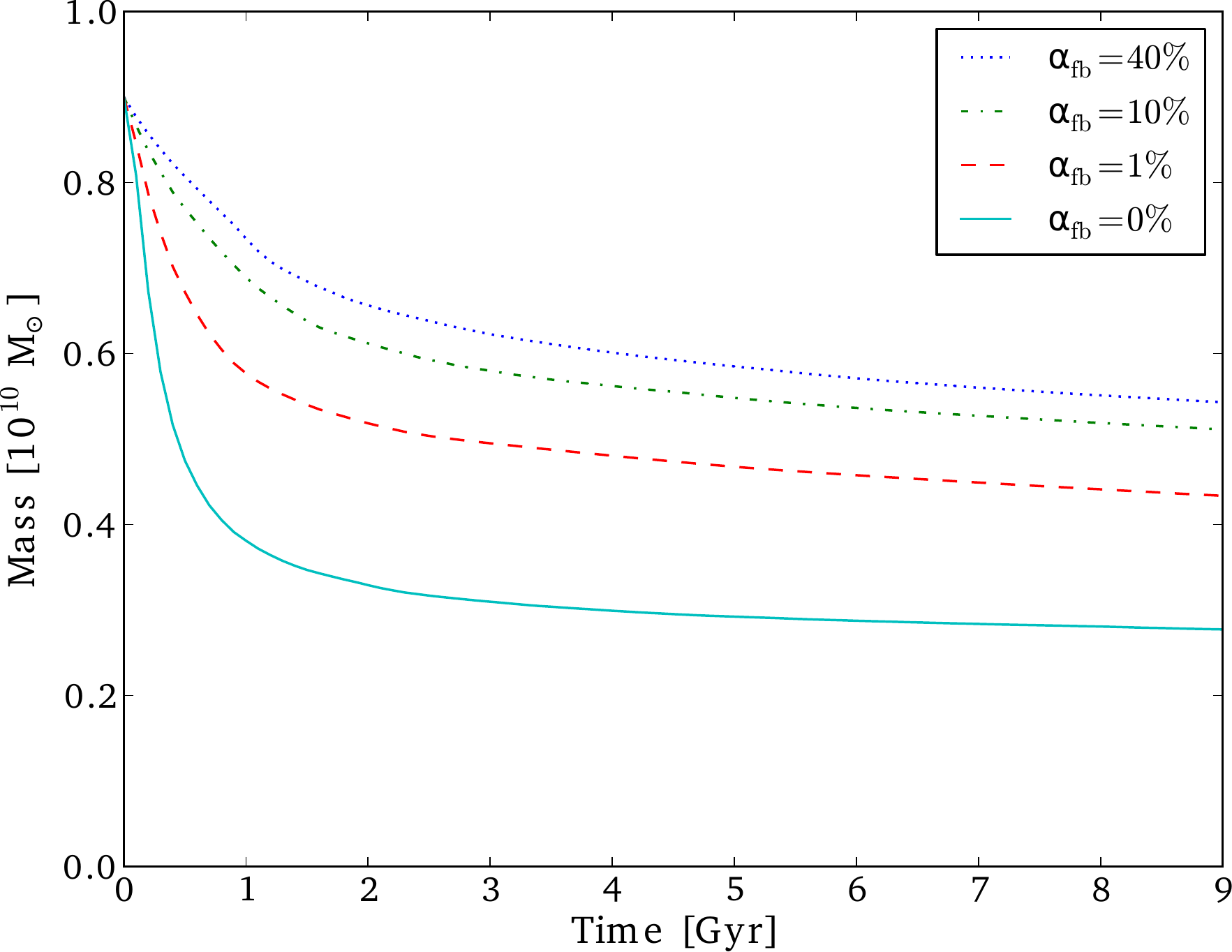}}
\caption{Time evolution of the mass of the gas component. Solid (blue) line: run without feedback. Dashed (green) line: run with a feedback efficiency $\alpha_{\mathrm{fb}}$=1$\%$. Dash-dotted (red) line: run with $\alpha_{\mathrm{fb}}$=10$\%$. Dotted (black) line: run with $\alpha_{\mathrm{fb}}$=40$\%$}
\label{massnoH2-fig}
\end{figure}

We plot in Figure~\ref{massnoH2-fig} the time evolution of the total mass of gas present in the simulations. All the gas is originally in the disc but can leave it under the effect of gravitational heating or stellar feedback. The characteristic time of consumption of the gas increases with feedback efficiency because of the moderating effect of feedback on gas density, regulating star formation, and the curves have increasing horizontal asymptotes y-values. This is due to gas expelled from the disc, and also to the inability of the gas to reach the star formation threshold. The gas in the outer parts of the disc remains diffuse with almost no star formation in all cases, as the lower abundance of metals does not allow the gas to cool down enough to form stars. More details concerning the star formation will be presented in the following section when compared to simulations including molecular hydrogen. 

\subsection{Simulations with H$_2$}
\label{H2-sec}
We now study the impact of the inclusion of H$_2$ on the gas physical state, star formation and structure of the discs.

\subsubsection{H$_2$ fraction}

\begin{figure}[!h]
\centering
\resizebox{\hsize}{!}{\includegraphics{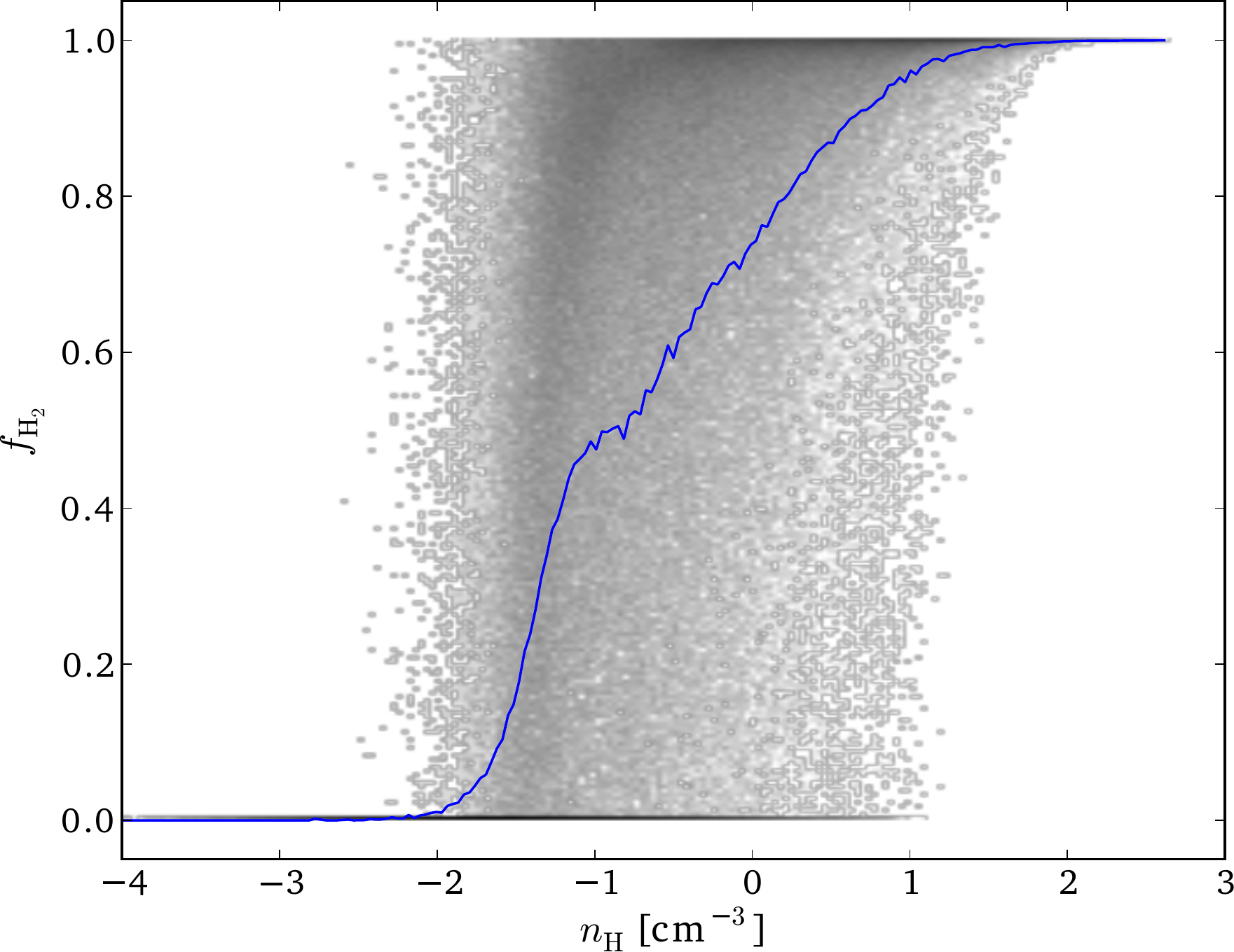}}
\caption{Mass fraction of H$_2$ versus hydrogen nuclei number density after 0.5~Gyr for $\alpha_{\mathrm{fb}}=1 \%$. The mass weighted 2D histogram is in grey scale while the mass-weighted 1D histogram is represented as the blue solid line.}
\label{fH2nH-fig}
\end{figure}

\begin{figure}[!h]
\centering
\resizebox{\hsize}{!}{\includegraphics{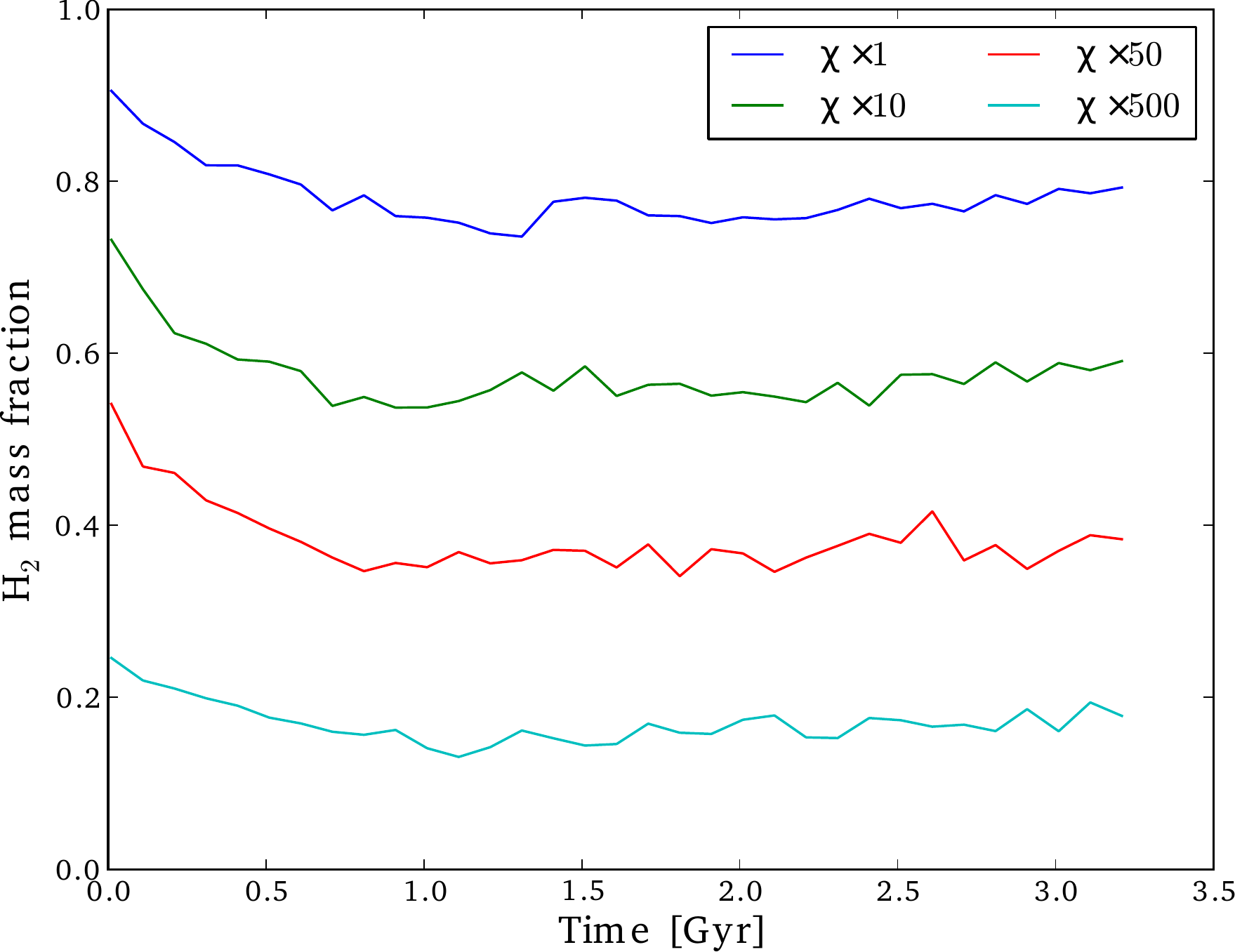}}
\caption{Global H$_2$ mass fraction versus time for $\alpha_{\mathrm{fb}}=1 \%$. The UV flux increases from top to bottom.}
\label{H2mass-fig}
\end{figure}

\begin{figure}[!h]
\centering
\resizebox{\hsize}{!}{\includegraphics{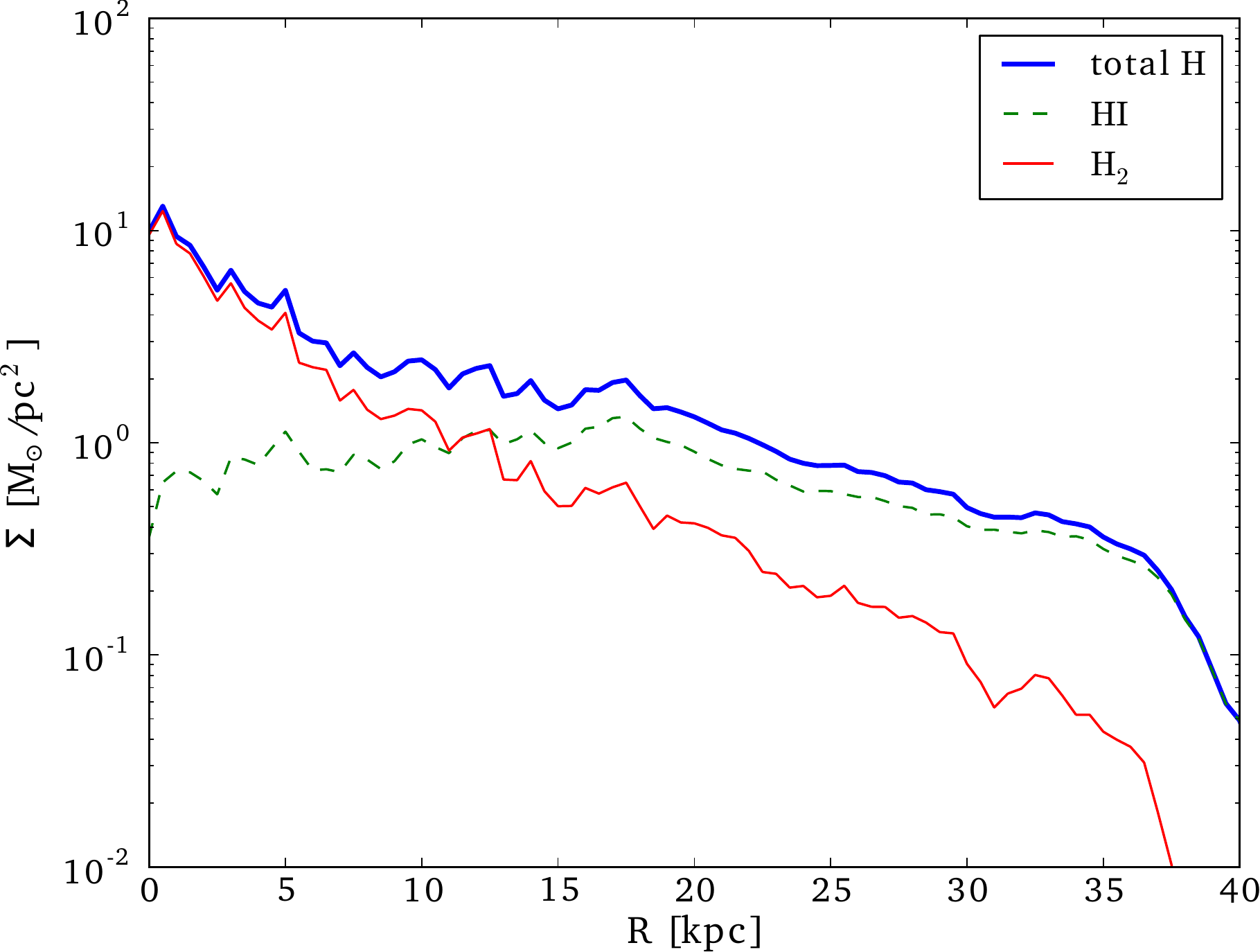}}
\caption{Radial distribution of the surface density of H$_2$, atomic gas HI and total hydrogen gas after 0.5~Gyr for $\alpha_{\mathrm{fb}}=1 \%$ and the selected UV scaling factor.}
\label{H2ofR-fig}
\end{figure}

We first set the feedback efficiency to $\alpha_{\mathrm{fb}}=1 \%$ and tune the $\chi$ factor in the H$_2$ fraction so that we find a global H$_2$ fraction that is consistent with observations while having  enough molecular hydrogen fraction to study its effects. Our metallicity gradient implies that H$_2$ will tend to be less present in the outer parts of the galaxy where the metallicity and dust abundance are low and the gas is less shielded from Lyman-Werner radiation. However, this is compensated by the lower star formation rate in these regions, which decreases the ambient Lyman-Werner luminosity. We plot the evolution of the total mass fraction of H$_2$ in Figure~\ref{H2mass-fig} for four different UV flux scaling factors. 
The exact exposition of the gas to UV flux is not well known, since it depends on the physics at very small scale, below our resolution, and on complex radiation transfer, through gas clumps and associated dust.
The global mass fraction first decreases with time because new stars are formed and contribute to the dissociating radiation field and becomes stable when the SFR becomes almost null (see the solid line of Figure~\ref{massH2-fig} for the evolution of the gas mass and the SFR). Figure~\ref{H2ofR-fig} shows the surface density of H$_2$ and atomic hydrogen gas HI after 0.5 Gyr of evolution for the case $\chi \times 50$ that we choose. Such a surface density profile is similar to some observed profiles of local galaxies described in \citet{young91}, and more recently in \citet{bigiel08}. 

We plot on Figure~\ref{fH2nH-fig} the mass fraction of H$_2$ versus the hydrogen nuclei number density at t=0.5~Gyr. The H$_2$ mass fraction is null for diffuse gas and equal to unity for the densest gas while the transition presents some scatter. This is due to the variation in the structure of the gas that reflects in the local average column density used in the method, and the variation in the amount of nearby recent star formation. The transition density is low compared to similar work by \citet{gnedin09} or \citet{chris121}. This is due to our calibration to obtain a realistic global H$_2$ mass fraction and our generally lower densities compared to \citet{gnedin09} and \citet{chris121}, arising partly from our choice of lower threshold density for star formation that prevents the gas from becoming as dense. For simulations with a higher threshold density for star formation, the calibration factor (that is varied on Figure~\ref{H2mass-fig}) should be higher.

\begin{figure}[!h]
\centering
\resizebox{\hsize}{!}{\includegraphics{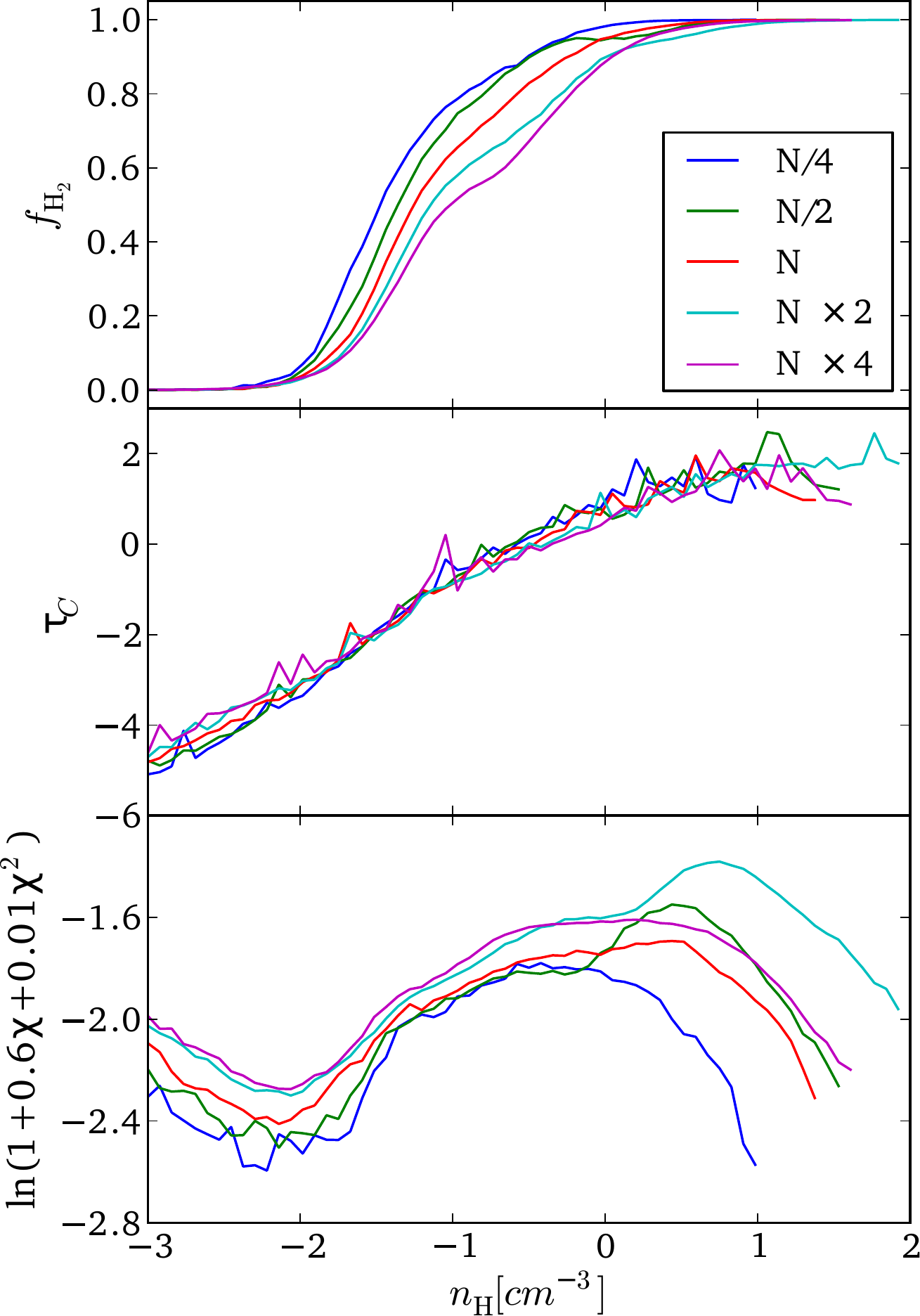}}
\caption{Different resolutions with $\alpha_{\mathrm{fb}}=10 \%$ after 0.5~Gyr of evolution. Top: H$_2$ mass fraction as a function of hydrogen nuclei number density. Middle: Dust optical depth as a function of hydrogen nuclei number density. Bottom: factor involving the UV flux entering the computation of the H$_2$ mass fraction (see equations~\ref{fh2-eq} and \ref{s-eq}) as a function of hydrogen nuclei number density.}
\label{res-fig}
\end{figure}

This method can be sensitive to the resolution of the simulation through the dependence of the molecular mass fraction on the column density proportional to $\dfrac{\rho^2}{|\nabla \rho|}$, and the $\chi$ factor proportional to $\dfrac{1}{\rho}$. We have run simulations with a feedback efficiency of $10 \%$ and varying number of particles: our reference $N_{\rm ref}$=1~200~000 and simulations with $\dfrac{N_{\rm ref}}{4}$, $\dfrac{N_{\rm ref}}{2}$, $N_{\rm ref} \times 2$ and $N_{\rm ref} \times 4$. The softening length $\epsilon$ has been modified respecting $\epsilon \propto N^{-1/3}$ with $N$ the number of particles, so that it is proportional to the average inter-particle distance. The other parameters, including the star formation efficiency, have been kept constant. The comparison between the simulations is delicate because simulations of different resolutions have a different density structures evolution, impacting the dust optical depth, and a different star formation history, impacting the $\chi$ factor. We represent the dust optical depth as a function of hydrogen nuclei number density for the different resolutions at the same simulation time on Figure~\ref{res-fig}. It shows practically no dependence on the resolution. The slight difference in the transition density between mostly atomic and mostly molecular hydrogen when resolution is changed is due to the higher average UV field as can be seen on the plot representing the distribution of the factor entering the computation of the molecular mass fraction $\ln (1+0.6\chi+0.01\chi^2)$ versus density.

\subsubsection{Gas physical state}
\label{gasphys-sec}
\begin{figure*}
\centering
\includegraphics[width=17cm]{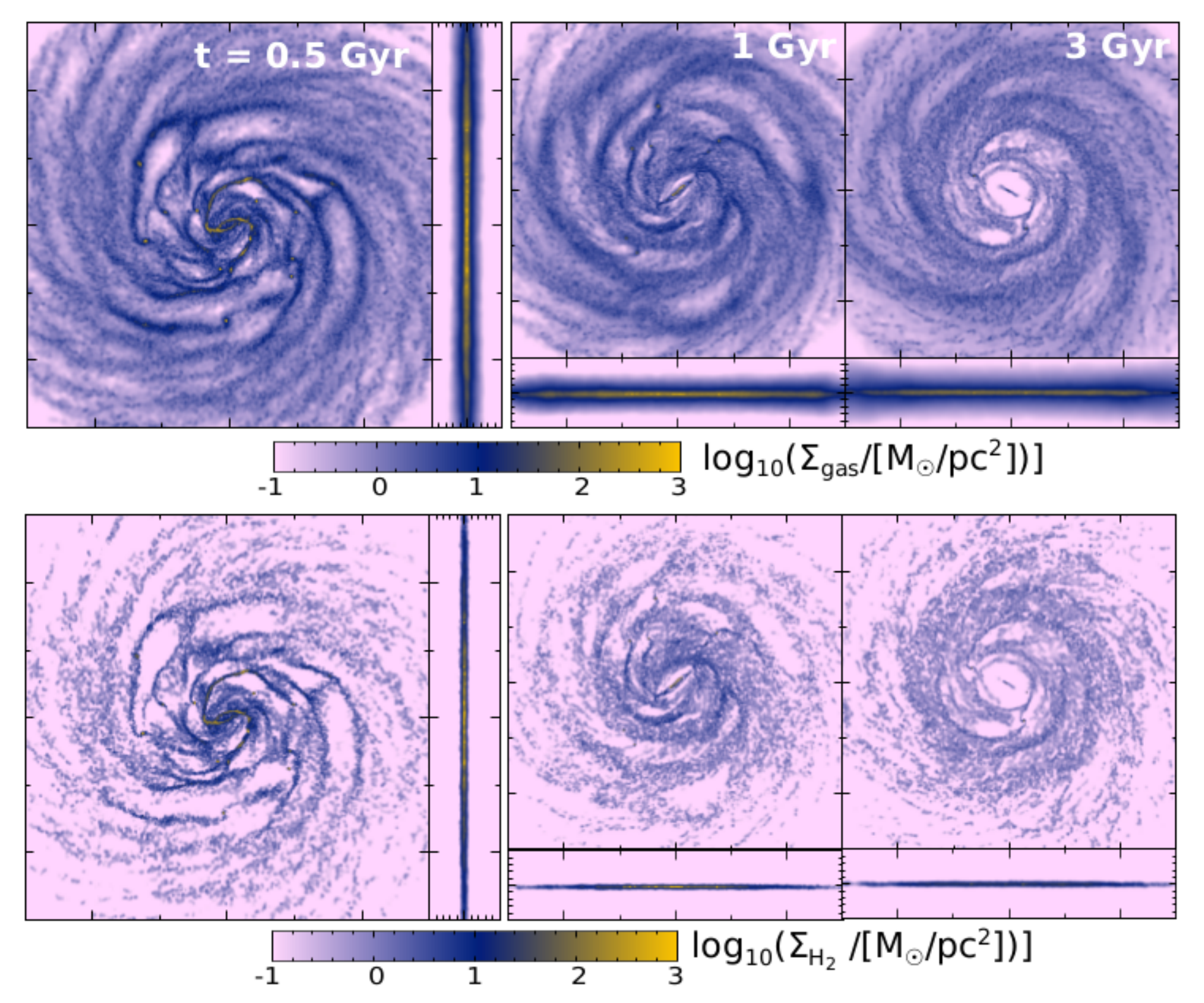}
\caption{Projections of the gas density (top row) and H$_2$ density (bottom row) after 0.5 Gyr, 1 Gyr and 3 Gyr of evolution. Box sizes are [60 kpc x 60 kpc] for face-on views and [60 kpc x 10 kpc] for edge-on views. The column integrated density scale is the same for all plots. Done with the visualisation software SPLASH \citep{price07}.}
\label{H2chiaspect-fig}
\end{figure*}

\begin{figure}[!h]
\centering
\resizebox{\hsize}{!}{\includegraphics{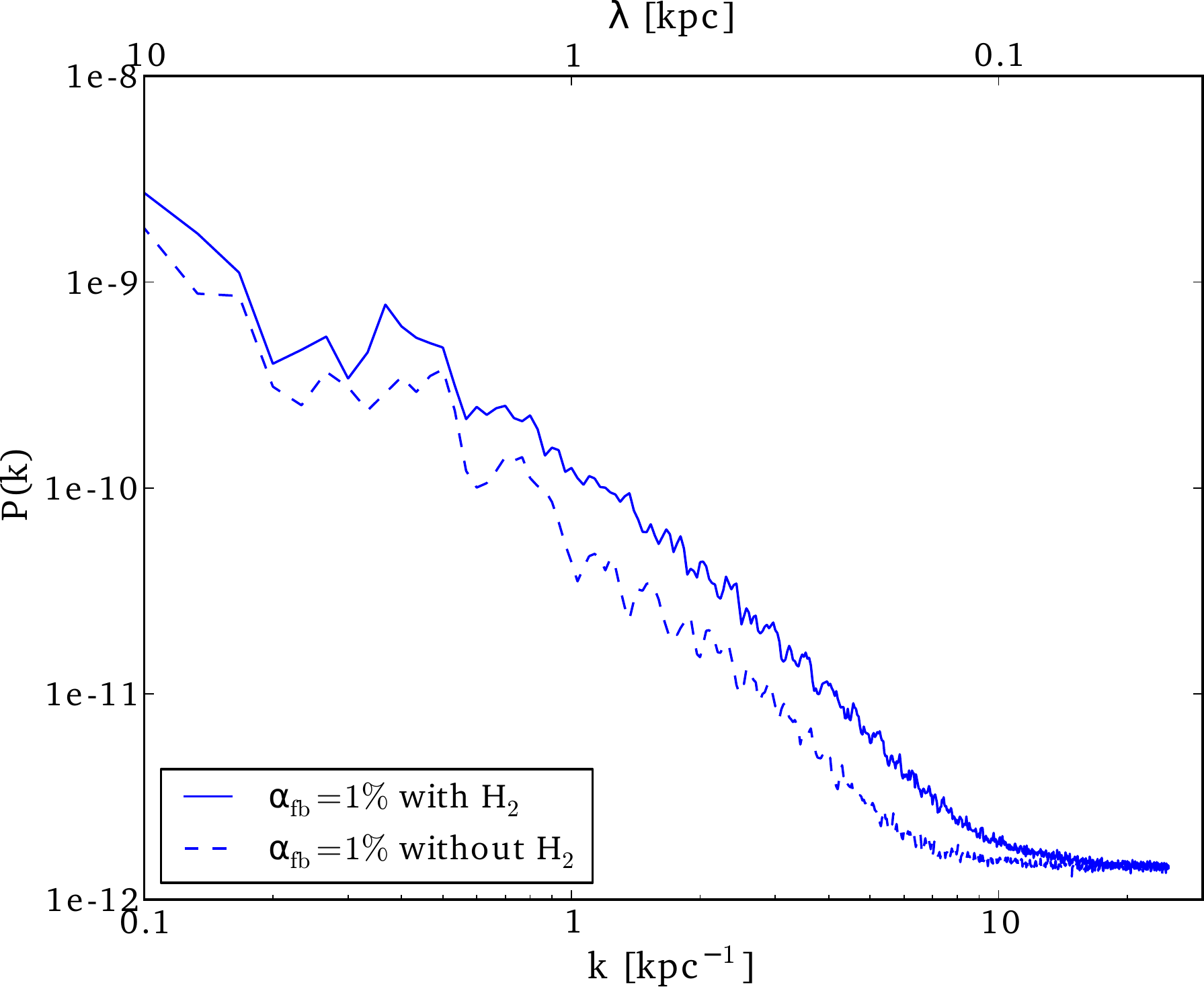}}
\caption{Gas surface density power spectra after 0.5~Gyr of evolution.}
\label{ps-fig}
\end{figure}

\begin{figure}[!h]
\centering
\resizebox{\hsize}{!}{\includegraphics{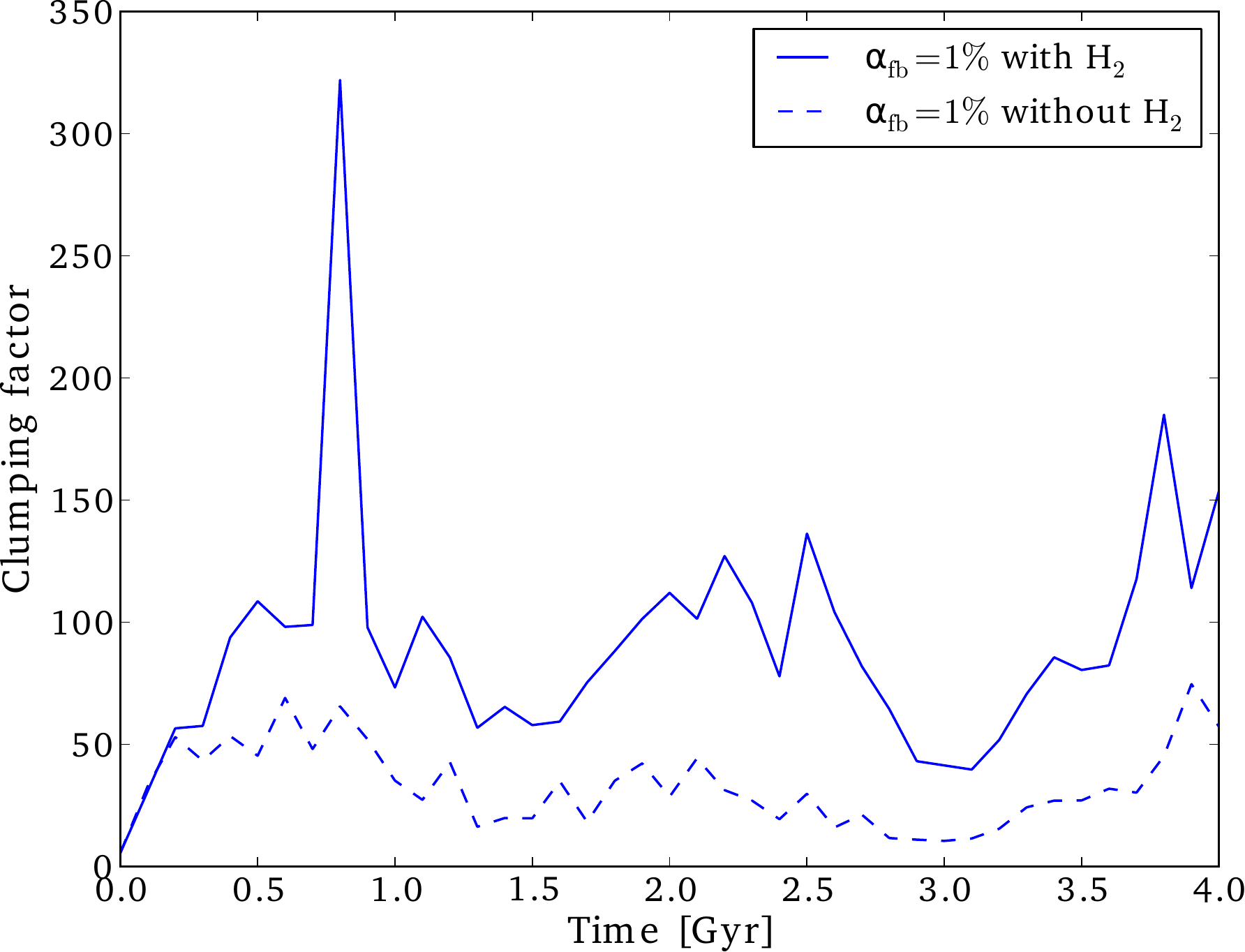}}
\caption{Clumping factor (defined in equation~\ref{clump-eq}) evolution with time.}
\label{clump-fig}
\end{figure}

Figure~\ref{H2chiaspect-fig} shows the aspect of the disc for $\alpha_{\mathrm{fb}}=1 \%$ at the same simulation epochs than in 
Figure~\ref{aspectnoH2-fig}, whose second row is the simulation for the same feedback efficiency but without H$_2$. 
Cooling by collisions of atomic hydrogen with metals in the purely HI simulation or mainly by H$_2$ in the now H$_2$ dominated central part make this region similar in both simulations, the difference is in the outer parts of the galaxy for which metal-line cooling is poorly efficient because of our assumed metallicity gradient. In this case, the gas remains diffuse with no other cooling processes, but the inclusion of H$_2$ allows the gas to be clumpier: we see density features that were absent in the purely atomic simulation. The surface density of H$_2$ is also plotted. It indeed follows the density features of the gas: the clumps and filamentary structures can be seen in H$_2$. We represent the corresponding power spectra of the gas surface density on Figure~\ref{ps-fig} for total gas masses normalised to the same value. It can be seen that the disc with molecular hydrogen has more small and intermediate size structures. We also plot the clumping factor evolution with time on Figure~\ref{clump-fig}. We define this clumping factor, as in \citet{springel032}, by:
\begin{equation}
C=\dfrac{\sum_i m_i \rho_i \sum_j m_j \frac{1}{\rho_j}}{\left(\sum_k m_k\right)^2}
\label{clump-eq}
\end{equation}

Figure~\ref{clump-fig} shows molecular hydrogen makes the gas clumpier.    

\begin{figure}
\centering
\includegraphics[width=7.5cm]{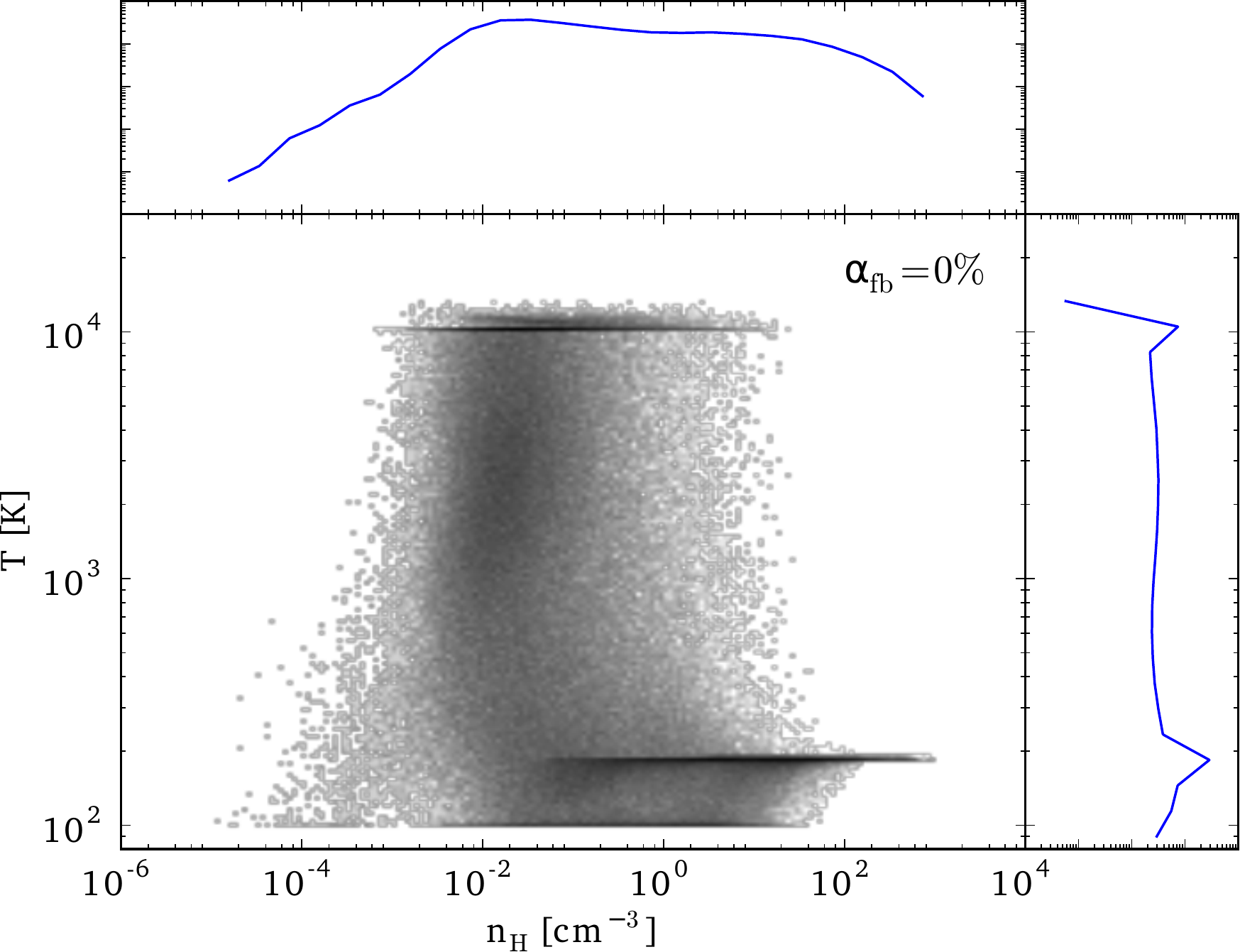}
\includegraphics[width=7.5cm]{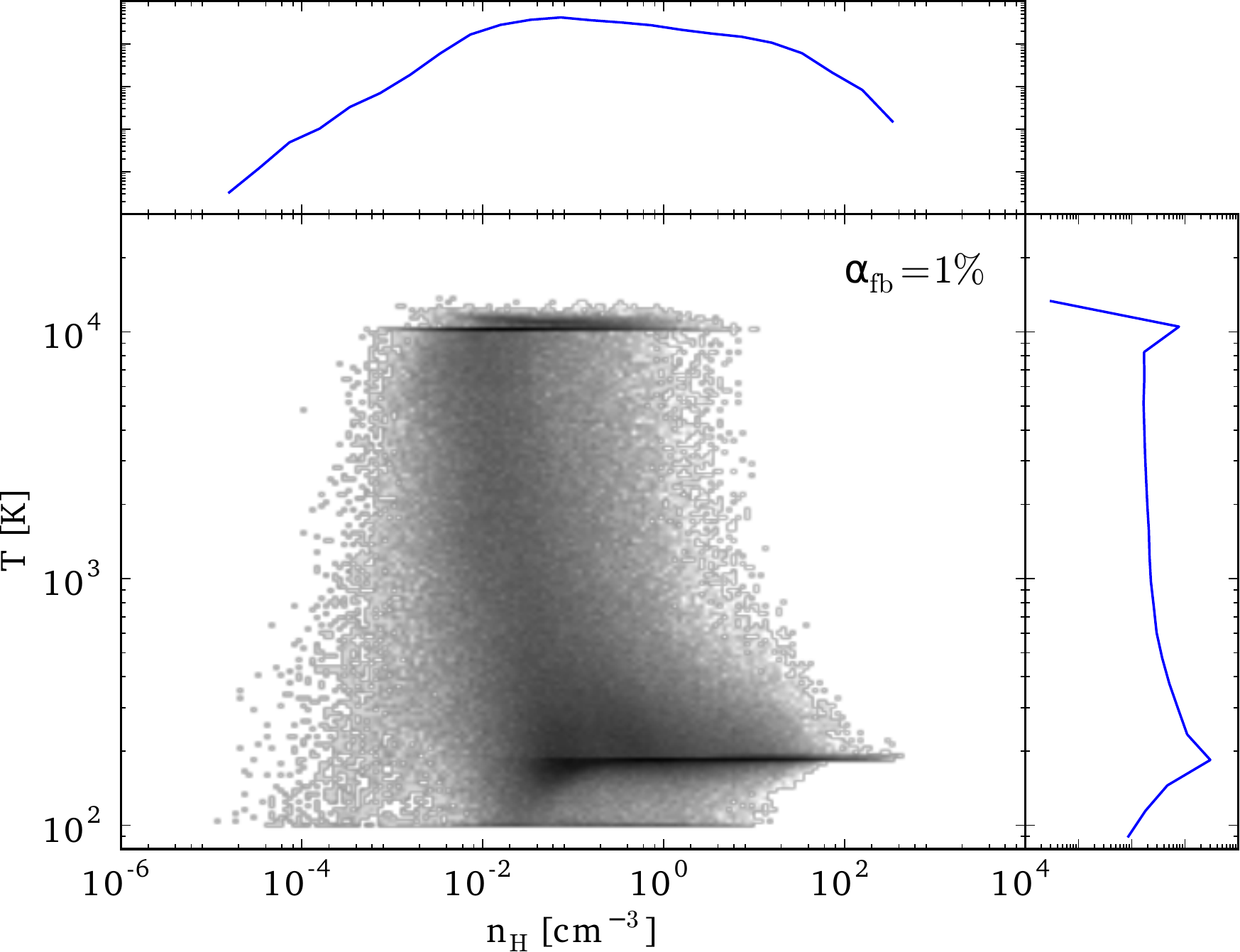}
\includegraphics[width=7.5cm]{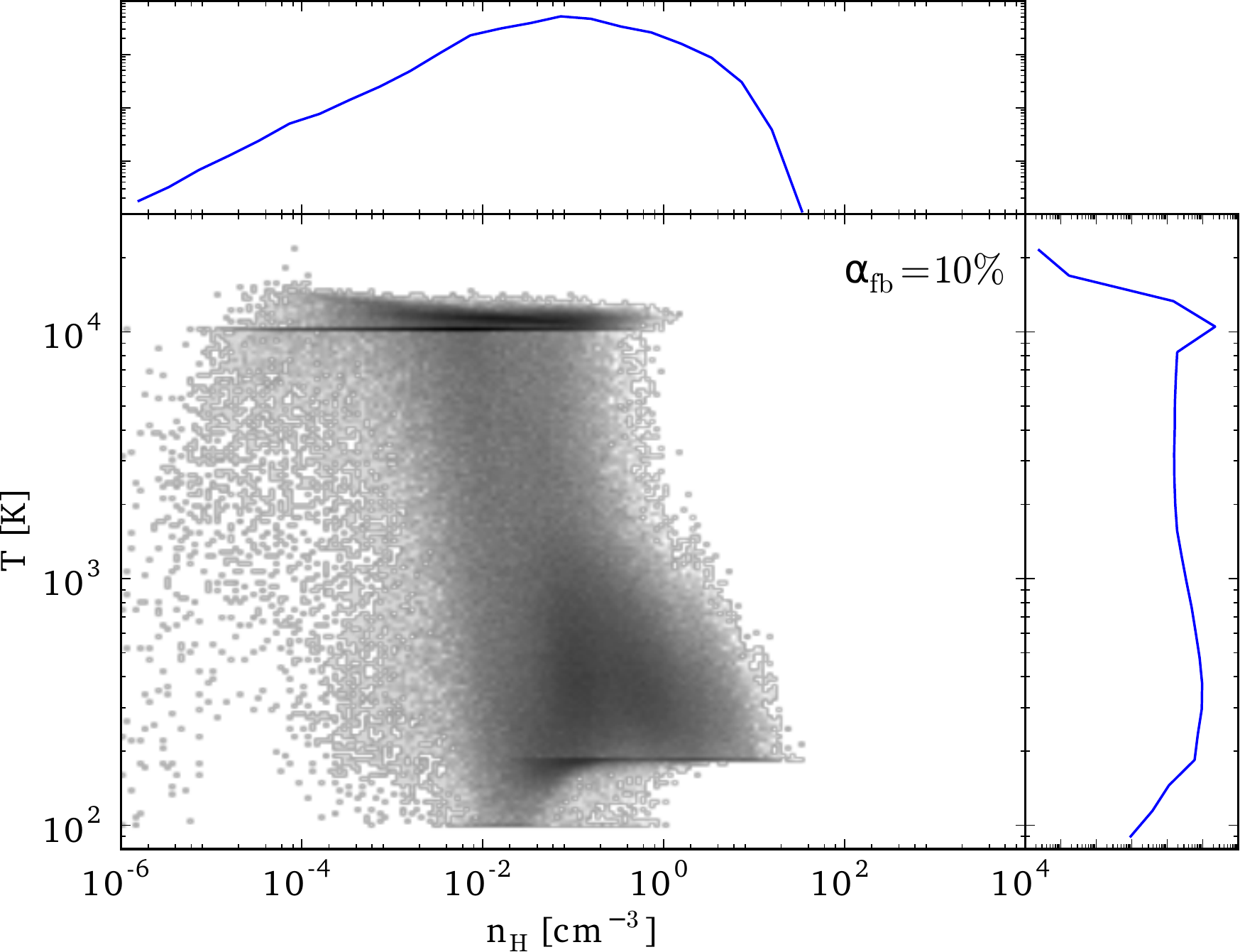}
\includegraphics[width=7.5cm]{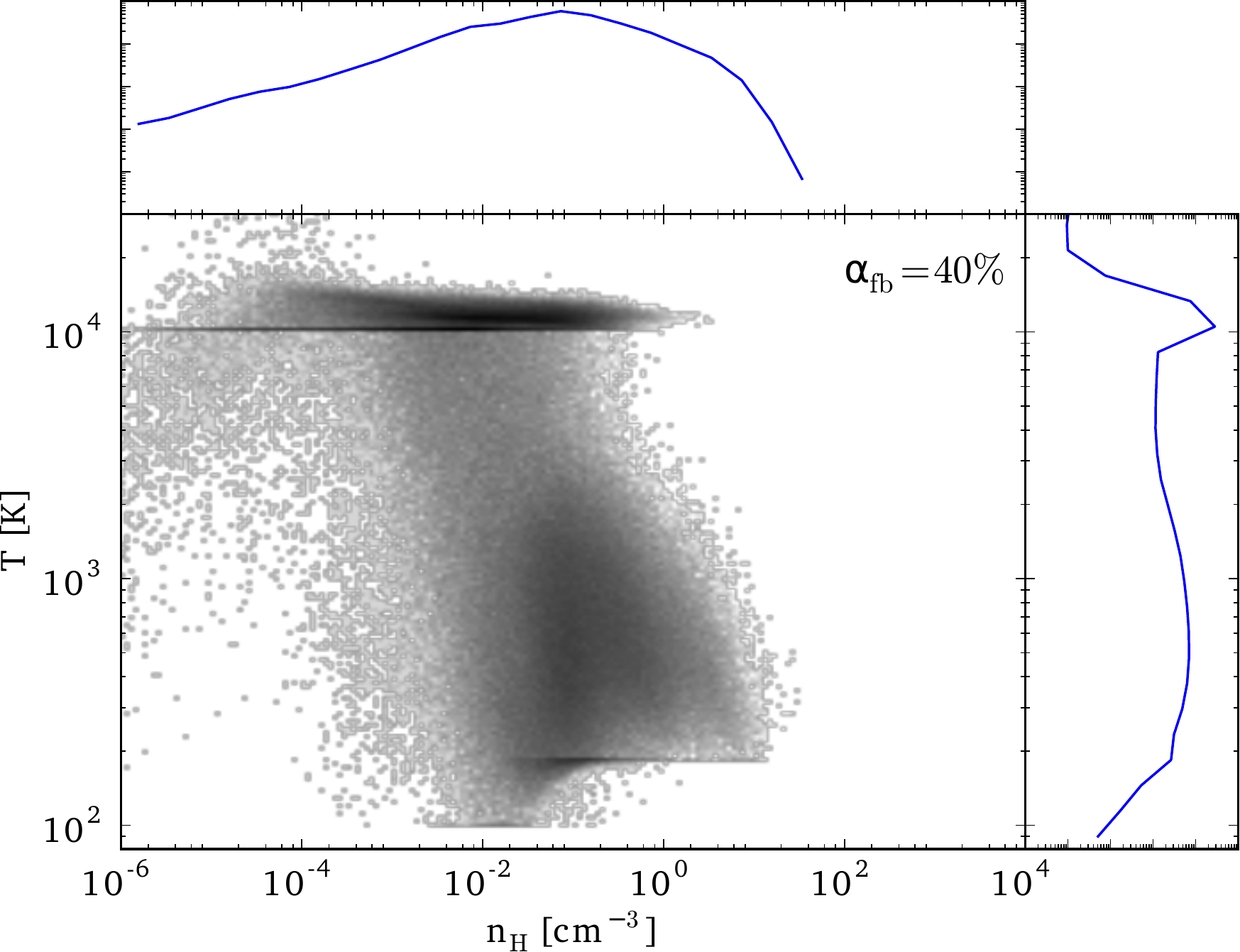}
\caption{Temperature-hydrogen number density histogram of the gas after 0.5 Gyr of evolution. The 2D histograms and marginal 1D histograms are all mass weighted and normalised. From top to bottom: runs with an increasing feedback efficiency.}
\label{hrTH2-fig}
\end{figure}

\begin{figure}[!h]
\centering
\resizebox{\hsize}{!}{\includegraphics{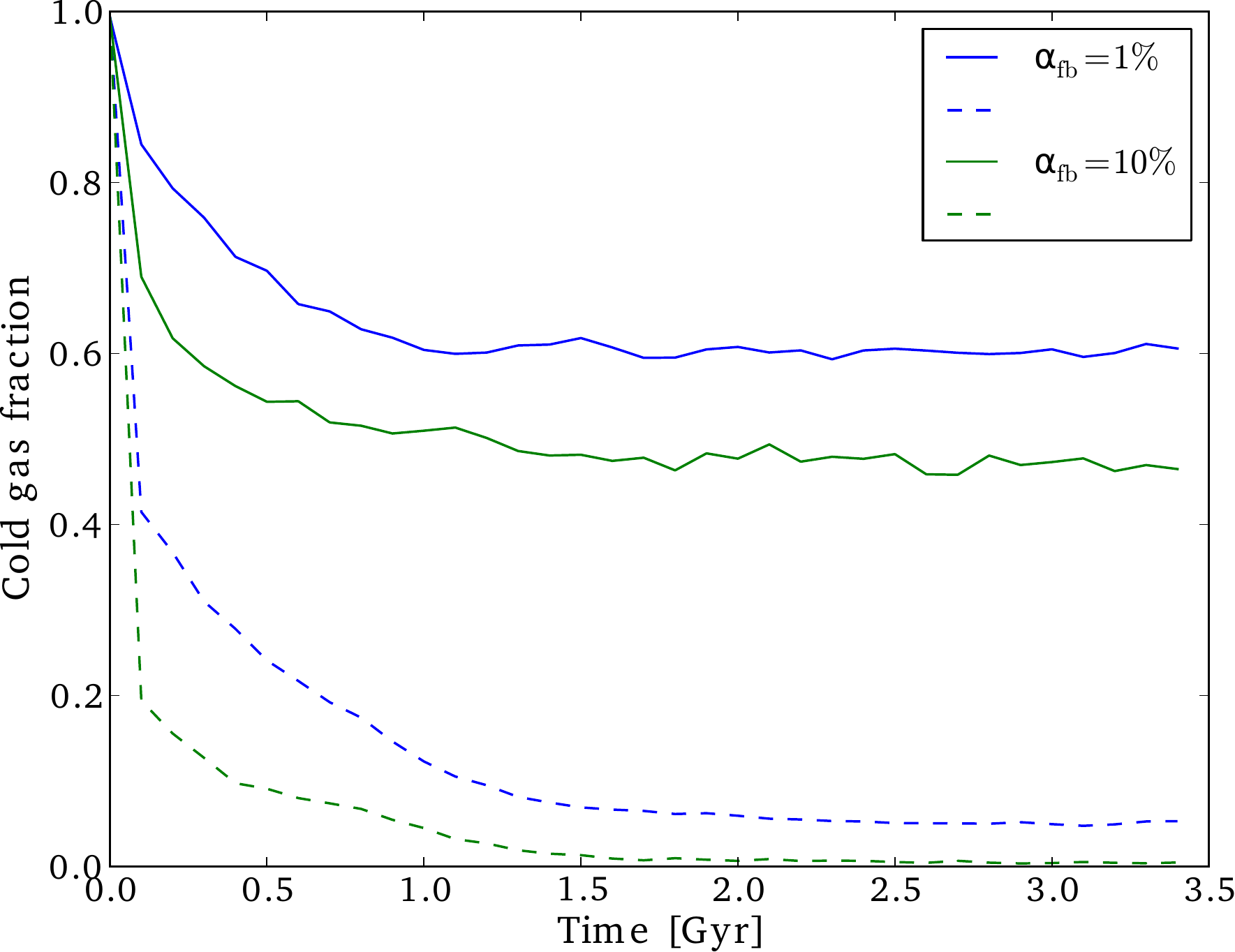}}
\caption{Fraction of cold gas as a function of time for two different feedback efficiencies  $\alpha_{\mathrm{fb}} = 1 \%$ and  $\alpha_{\mathrm{fb}} = 10 \%$. Solid lines: with H$_2$. Dashed lines: without H$_2$.}
\label{coldwarm-fig}
\end{figure}
The temperature-number density histograms of Figure~\ref{hrTH2-fig} and the marginal PDFs show higher fractions of gas in the cold dense phase than for the corresponding feedback efficiencies without molecular hydrogen. There is no clear diagonal branch because there is some efficient cooling in the whole disc. We separate the gas in two ranges of temperature, below and above 1000~K. We study the evolution of the fractions of cold and warm gas (gas lying below or above this threshold) depending on the feedback efficiency, and with or without the inclusion of cooling by molecular hydrogen. The majority of star formation happens in the first Gyrs in all the simulations (especially when the feedback efficiency is low and stars form quickly), so we focus on this period and plot the cold gas fraction as a function of time on Figure~\ref{coldwarm-fig} for two feedback efficiencies. All the gas is initially in the cold phase. For a given feedback efficiency, the cold gas phase represents a much higher fraction of the gas if H$_2$ cooling is taken into account. Without H$_2$, stars are formed from warmer and more diffuse gas, reducing the star formation efficiency.
Feedback decreases the fraction of cold gas: the kinetic energy given to particles in dense star forming regions is transformed into thermal energy by pressure forces and viscosity, all the more as the feedback efficiency is high. 

\begin{figure}[!h]
\centering
\resizebox{\hsize}{!}{\includegraphics{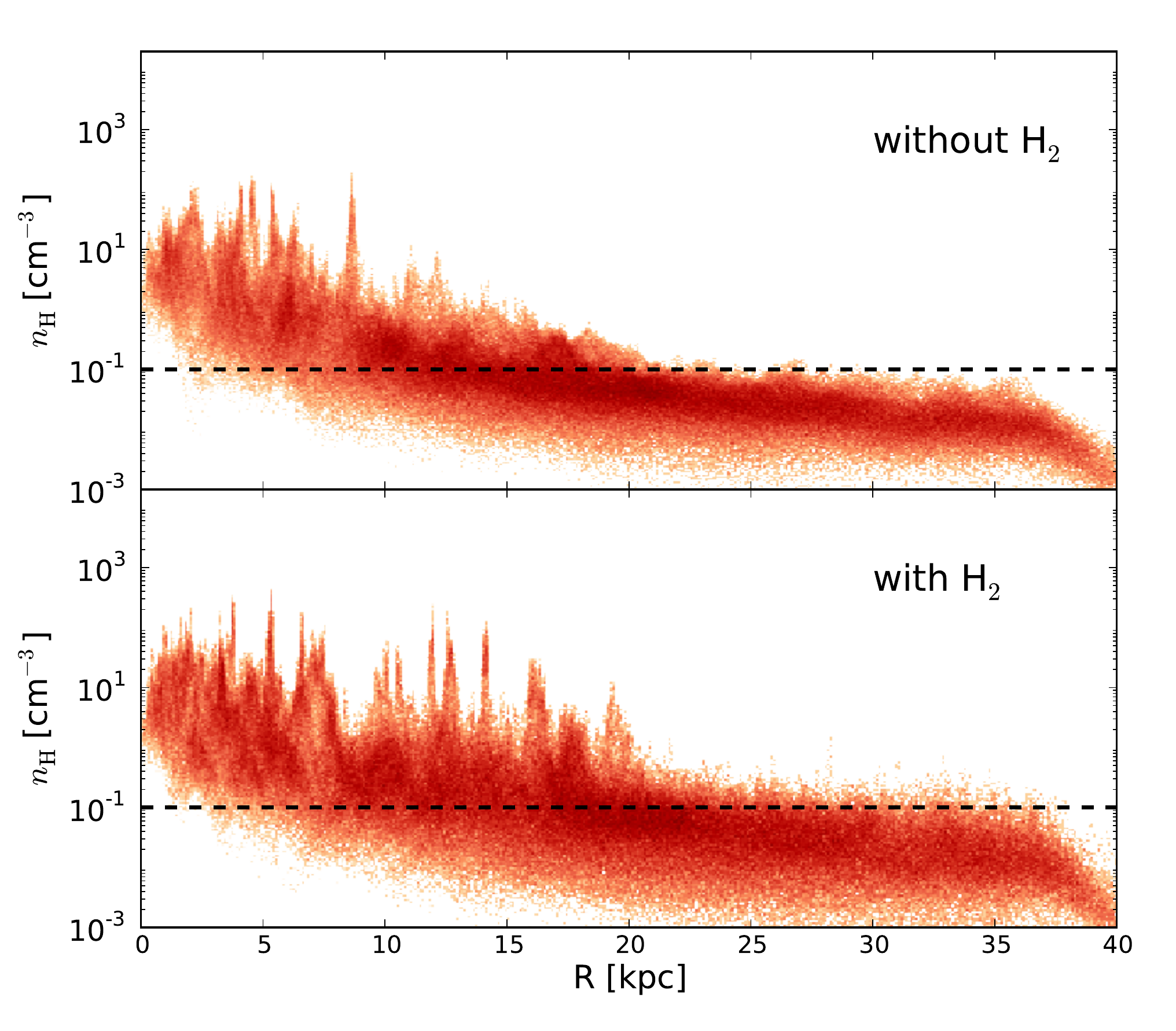}}
\caption{PDFs of hydrogen number density as a function of galactocentric radius after 0.5 Gyr of evolution with or without including H$_2$ cooling in the simulations, for $\alpha_{\mathrm{fb}} = 1 \%$. Horizontal black line: threshold density for star formation.}
\label{nHofR-fig}
\end{figure}

As we are especially interested in studying the state of the gas as a function of galactocentric radius, we plot the histograms of gas density versus radius in Figure~\ref{nHofR-fig} for the atomic and molecular simulations having $\alpha_{\mathrm{fb}}=1 \%$ , after 0.5~Gyr of evolution. The effect on density is clear: the gas exhibits density peaks at larger radii and has a mean higher density with the inclusion of H$_2$, with gas denser than the star formation threshold even at large radii. 

\begin{figure}[!h]
\centering
\resizebox{\hsize}{!}{\includegraphics{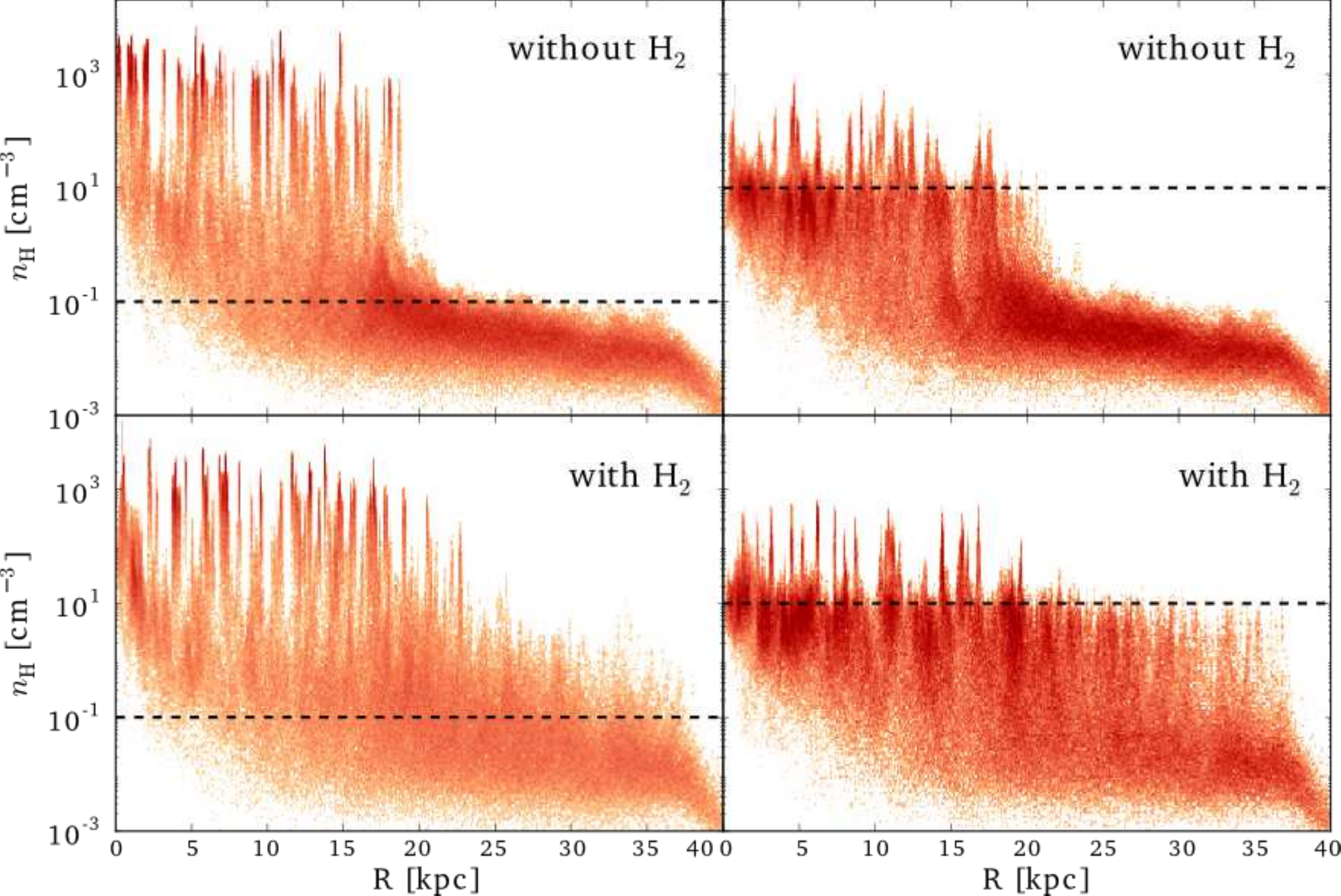}}
\caption{PDFs of hydrogen number density as a function of galactocentric radius after 0.5 Gyr of evolution with or without including H$_2$ cooling in the simulations, for $\alpha_{\mathrm{fb}} = 1 \%$ and alternative star formation parameters. Left: runs with $n_{\rm H_{min}}=10^{-1}$cm${-3}$ and $c_*=0.01$. Right: runs with $n_{\rm H_{min}}=10^1$cm${-3}$ and $c_*=0.1$. Horizontal black line: threshold density for star formation.}
\label{nHofRalt-fig}
\end{figure}

The interstellar gas we obtain is heterogeneous but could be even more so with different parameters for star formation. We also run simulations with a lower star formation efficiency per free-fall time $c_*=0.01$ and the same threshold density $n_{\rm H_{min}}=10^{-1}$cm${-3}$, and simulations with the same $c_*=0.1$ but a higher $n_{\rm H_{min}}=10^{1}$cm${-3}$. Reducing $c_*$ or increasing $n_{\rm H_{min}}$ both allow for higher density peaks, as seen on Figure~\ref{nHofRalt-fig}. The effect of including H$_2$ is also visible in these sets of simulations, but the H$_2$ abundance is overestimated here as the density is on average higher than in our other simulations but we run these new simulations without changing the calibration of the $\chi$ factor in the molecular mass fraction. When c$_*$ is lowered or $n_{\rm H_{min}}$ is heightened, the gas forms stars more difficultly and can thus collapse more efficiently, resulting in a more fractioned gas.

\subsubsection{Star formation}

\begin{figure}[!h]
\centering
\resizebox{\hsize}{!}{\includegraphics{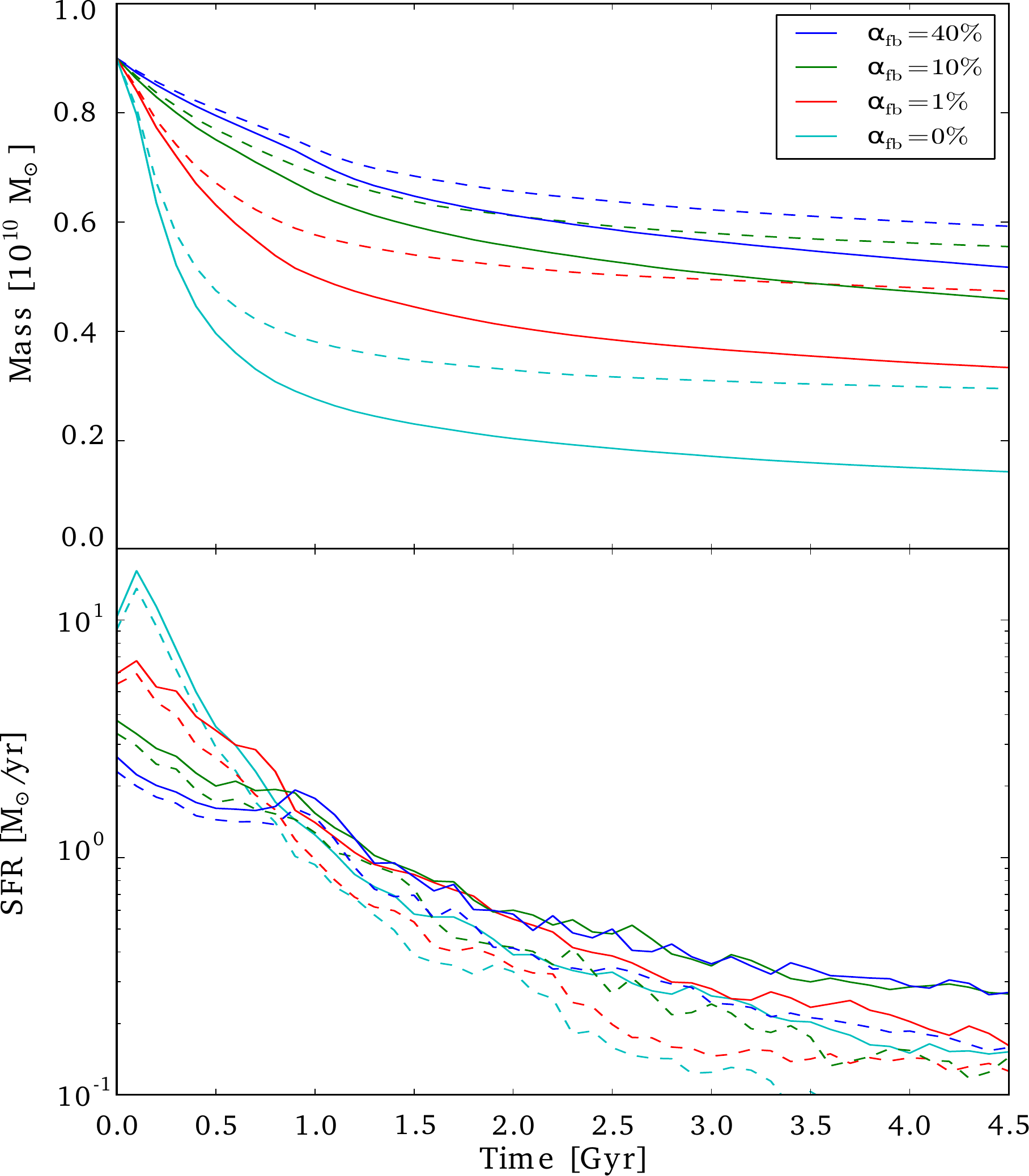}} 
\caption{Top: evolution of the gaseous mass of the galaxy. Bottom: SFR evolution. Solid lines: runs with H$_2$. Dashed lines: runs without H$_2$.}
\label{massH2-fig}
\end{figure}

The effect of H$_2$ cooling on the star formation efficiency is visible in Figure~\ref{massH2-fig}. Figure~\ref{massH2-fig} shows the total gas mass evolution with time and the SFRs for the different simulations. The depletion time of the gas decreases when molecular hydrogen cooling is included. 

We also show the SFRs as a function of time for our alternatives star formation parameters presented at the end of  section~\ref{gasphys-sec}. The reduced ability of the gas to form stars results in a delayed peak of star formation, after some time during which the gas has got denser and denser through cooling before being consumed by star formation.

\begin{figure}[!h]
\centering
\resizebox{\hsize}{!}{\includegraphics{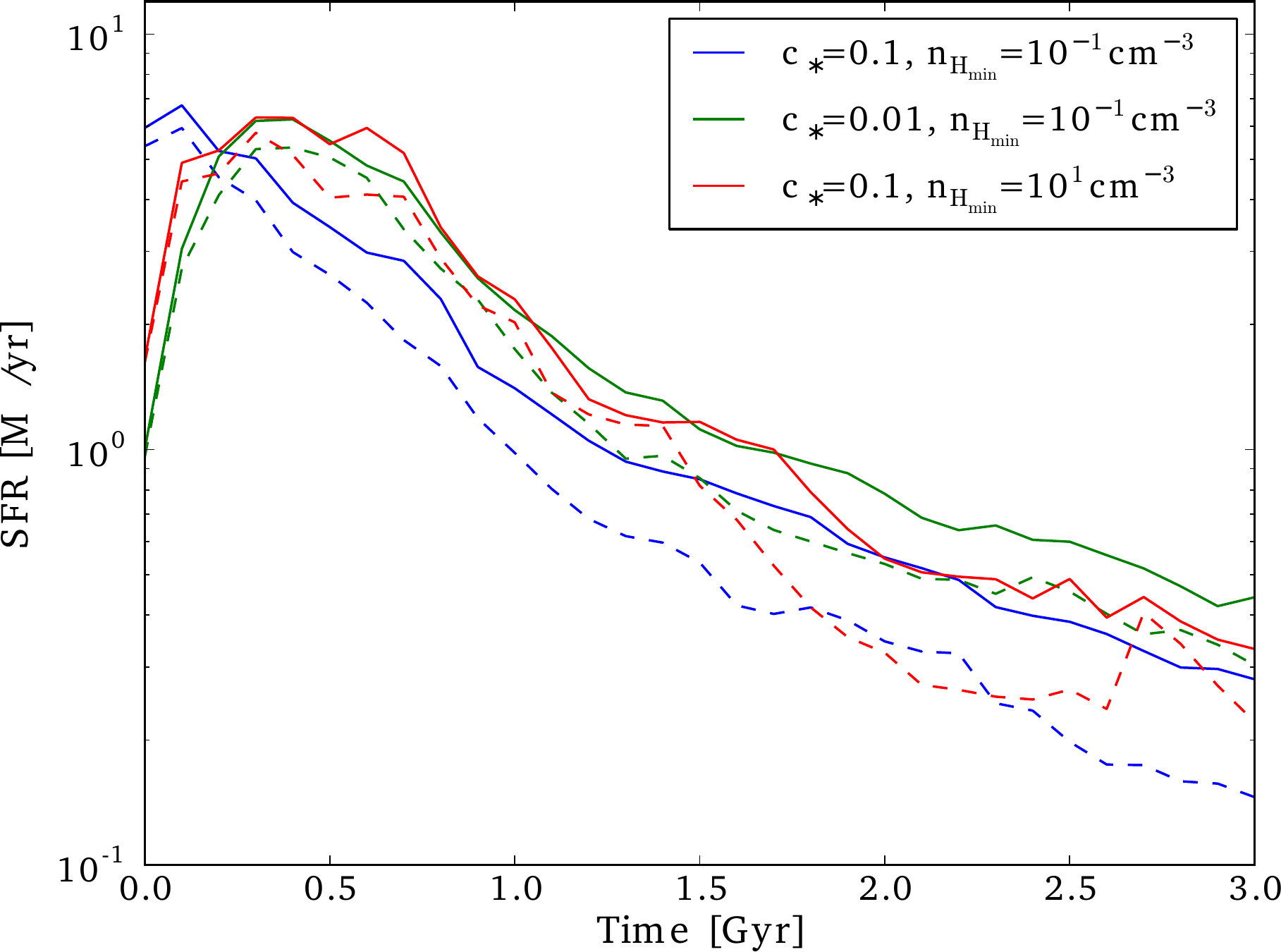}}
\caption{SFR evolution for $\alpha_{\mathrm{fb}} = 1 \%$ and alternative star formation parameters. Solid lines: runs with H$_2$. Dashed lines: runs without H$_2$.}
\label{sfralt-fig}
\end{figure}

\begin{figure}[!h]
\centering
\resizebox{\hsize}{!}{\includegraphics{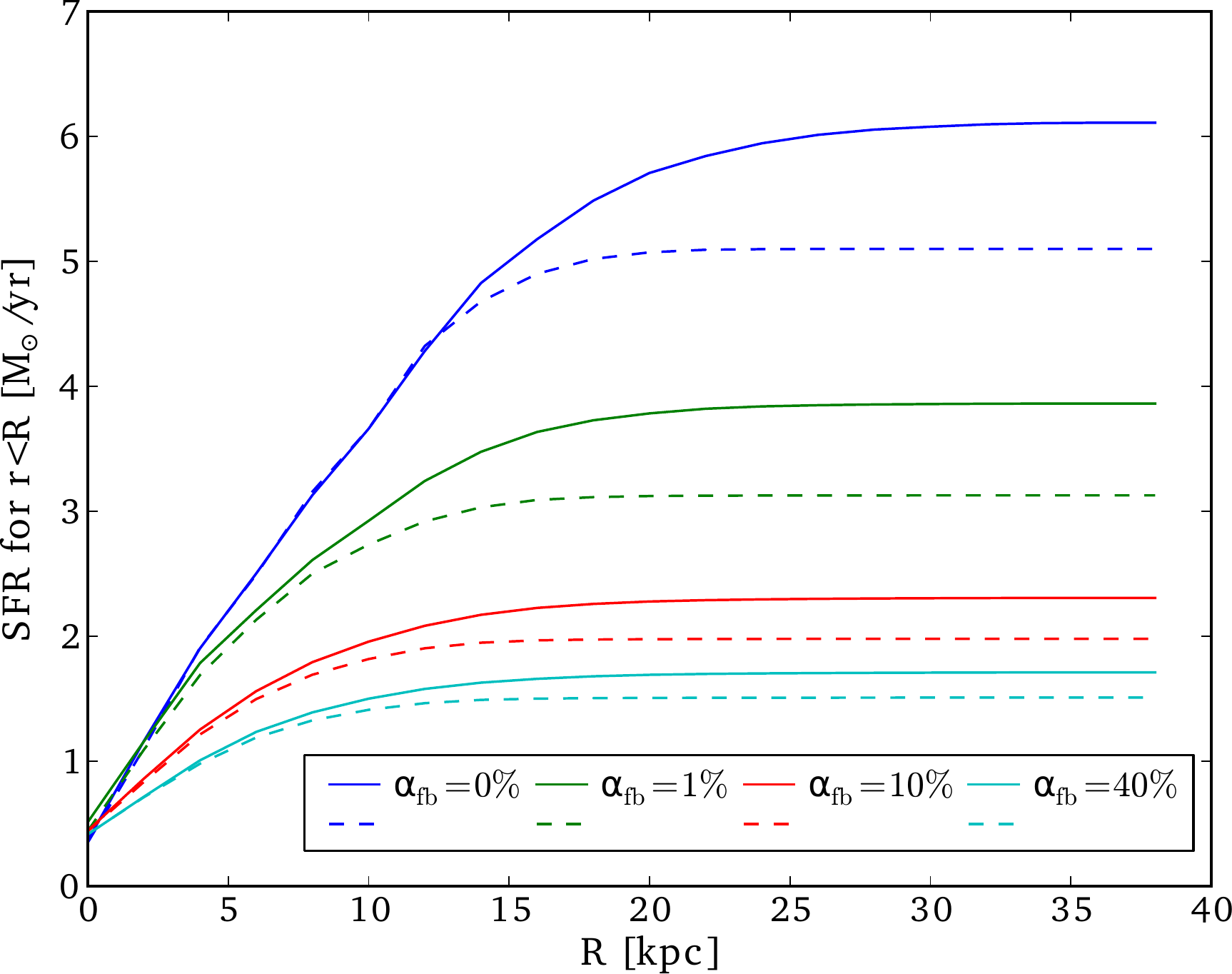}}
\caption{Cumulative star formation rate as a function of galactocentric radius, averaged on the first Gyr. Solid lines: run with H$_2$. Dashed  lines: run without H$_2$}
\label{sfrofR-fig}
\end{figure}

The star formation efficiency is similar in the central regions, but the inclusion of H$_2$ allows for star formation in the outer parts too. Having inserted the formation time of stars in our simulation outputs, we can track star formation spatially. Our outputs are temporally spaced by 10~Myr. We define the SFR as being the mass of stars formed during 10~Myr divided by this time, which is similar to the star formation rates obtained from observations in H$_{\alpha}$. We plot the cumulative SFR averaged on the first Gyr as a function of radius for these two simulations and other feedback efficiencies in Figure~\ref{sfrofR-fig}, and we indeed see that about the same amount of star formation occurs in the central parts, but molecular hydrogen starts playing a role at large radii. The difference occurs at a larger radius for the simulation with no feedback, which can be explained by the already high clumping in central parts of the disc, making star formation very efficient even without H$_2$. 

\begin{figure*}
\centering
\includegraphics[width=17cm]{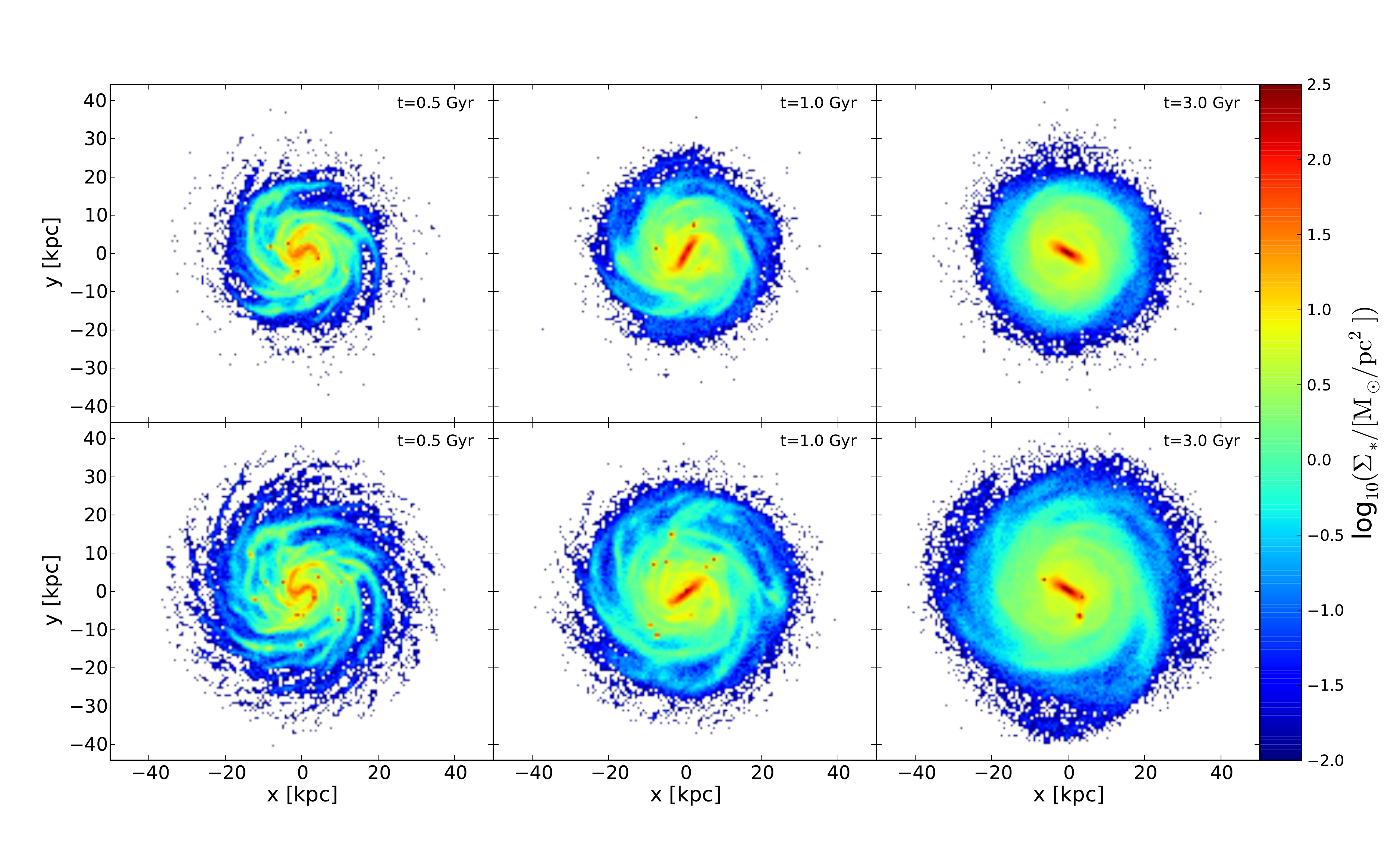}
\caption{Projection of the density of stars formed during the simulations after 0.5 Gyr, 1 Gyr and 3 Gyr of evolution. Top row: feedback efficiency $\alpha_{\mathrm{fb}} = 1 \%$ and no H$_2$. Bottom row: feedback efficiency $\alpha_{\mathrm{fb}} = 1 \%$ and H$_2$.}
\label{starssnap-fig}
\end{figure*}

Figure~\ref{starssnap-fig} shows the projected density of stars formed since the beginning of the simulation for $\alpha_{\mathrm{fb}}=1 \%$: the disc of new stars is more extended if we include H$_2$. There is a few stellar clumps, and also a very clear stellar bar that is maintained in the stellar component after a few Gyrs in both cases. Clumps of young stars can be seen on the density maps: they follow the gas clumps. These stellar clumps gradually lose energy and are eventually absorbed by the central bar.

We further study the star formation by drawing Kennicutt-Schmidt (KS) diagrams representing the surface density of SFR 
as a function of the gas surface density. As we are limited in mass resolution for star formation (stellar particles of a fixed mass of $\sim 10^4$ M$_{\odot}$ are created stochastically, completely differently from some smooth star formation), in order to have a significant amount of data to study, we add data points corresponding to 50 snapshots, from t=200~Myr to t=700~Myr. 
We use a polar grid with a given number n$_R$ of bins in cylindrical radius R and a given number n$_{\theta}$ of bins 
in azimuthal angle $\theta$. Using this kind of grid allows for a more uniform signal/noise repartition than with an orthogonal grid, and optimizes the number of new stars per cell. 

\begin{figure*}
\centering
\resizebox{\hsize}{!}{\includegraphics{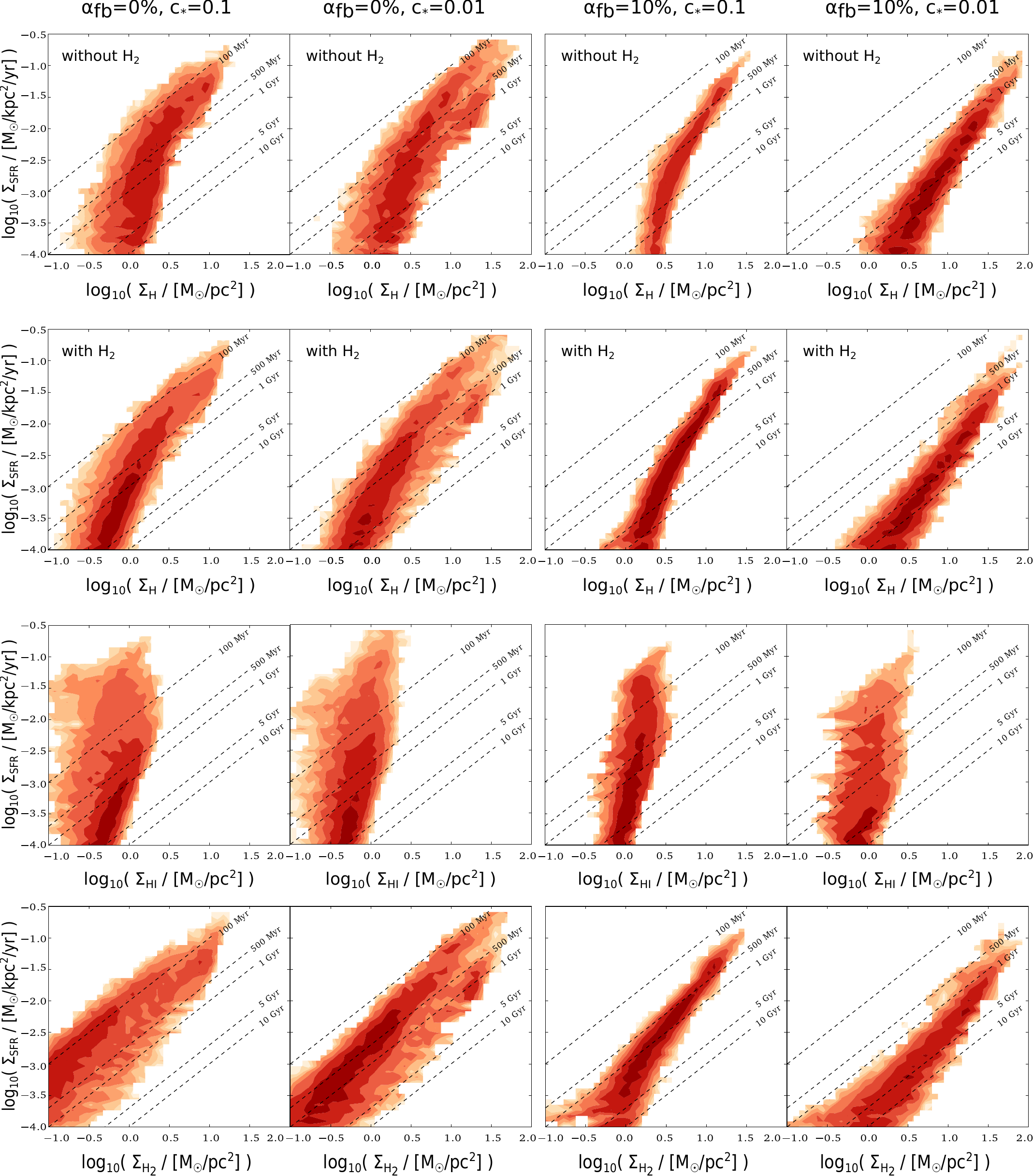}}
\caption{K-S diagrams. Two left columns: no feedback. Two right columns: $\alpha_{\mathrm{fb}}=10 \%$. Star formation efficiencies per free fall time $c_*$ are indicated on the top. Top row: simulations without H$_2$. Second row: simulations with H$_2$. Two bottom rows: SFR versus the HI only and H$_2$ only surface densities for simulations with H$_2$.}
\label{KS-fig}
\end{figure*}

On Figure~\ref{KS-fig}, we have plotted KS diagrams for simulations without feedback, and with a feedback 
efficiency $\alpha_{\mathrm{fb}} = 10 \%$. These can be compared with \citet{ager12} plots of azimuthally averaged KS diagrams of disc galaxies for different feedback intensities. Very similarly, we observe a global diminution of the SFR surface density 
when we increase the feedback strength. The figure quantifies how, on average, for the same gas surface densities, 
the SFR is lower with higher feedback. The feedback makes the gas more diffuse and destroys clumps.
 Two cells containing the same amount of gas but different fractions of diffuse gas (cells are larger than the clumps size), 
will have different star formation efficiency. This explains the smaller scatter in the simulations with feedback: 
as the gas is more homogeneous, the relation between surface densities of SFR and gas is better determined. 
In Figure~\ref{KS-fig}, lines of constant gas depletion time are indicated. The gas depletion time is defined as 
$t_{\mathrm{dep}}=\dfrac{\Sigma_{\mathrm{Gas}}}{\Sigma_{\mathrm{SFR}}}$. 
It can be seen that the high SFR and gas surface density regions of the galaxies have a depletion time as low as a 
few hundreds of Myrs in the simulations with no feedback, while the low SFR and density regions have depletion 
times up to 10~Gyr. The outer parts of the disc with a low surface density form stars much less efficiently
 than the central parts. We show simulations with an alternative lower star formation efficiency per free-fall time $c_*=0.01$. In this case, the KS diagrams are shifted towards the right as densities can be higher due to the higher clumping, and star formation is less efficient for a given volume density, which roughly translates in a given surface density. The scatter for $\alpha_{\mathrm{fb}} = 10 \%$ is a little higher for $c_*=0.01$ than $c_*=0.1$ because the gas is clumpier for a lower $c_*$ and the relation between volume and surface density departs there again from a one-to-one relation.

We also see a difference for low surface densities between only atomic and atomic and molecular simulations. 
The SFR is varying more linearly with molecular gas than with atomic gas. And the low surface density regions
form stars more efficiently with molecular gas.
This is explained by the fact that H$_2$ cooling allows the gas to be locally denser. 
It especially allows it to be more concentrated in the disc plane as can be checked in the edge-on projections.
 Then the gas is denser in volumic density, and forms stars more efficiently, at a given 
surface density. 

For simulations including H$_2$, the SFR is shown separately as a function of atomic and molecular components,
on the bottom of Figure~\ref{KS-fig}. The atomic hydrogen surface density is confined to low values for our galaxies 
and the SFR-HI diagrams show a large scatter because HI is too diffuse to track the star forming gas. The fact that 
 H$_2$ is a better tracer of star formation is found in observations of nearby galaxies (e.g. Bigiel et al 2008) and our results are in agreement with a number of theoretical work \citep[e.g.][]{maio10,glover08,greif10,wada09,chris122,gnedin09}.

\subsubsection{Vertical structure of the disc}

\begin{figure*}
\centering
\includegraphics[width=17cm]{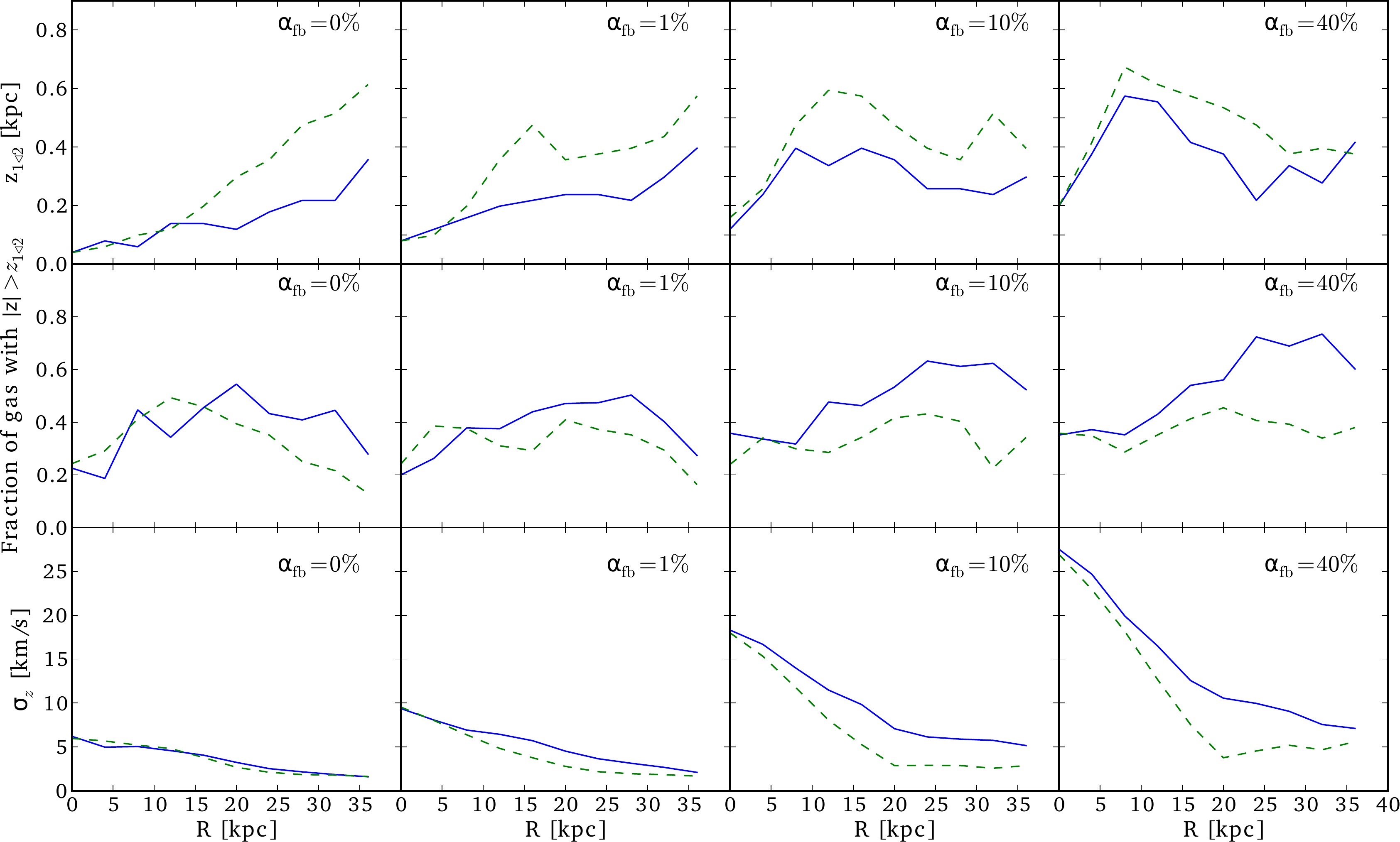}
\caption{Vertical structure of the gas. First row: characteristic gas height z$_{1/2}$ as a function of galactocentric radius after 0.5 Gyr of evolution. Second row: fraction of gas beyond the characteristic height. Third row: vertical velocity dispersion. Solid lines: runs with H$_2$. Dashed lines: runs without H$_2$}.
\label{zofR-fig}
\end{figure*}

\begin{figure}[!h]
\centering
\resizebox{\hsize}{!}{\includegraphics{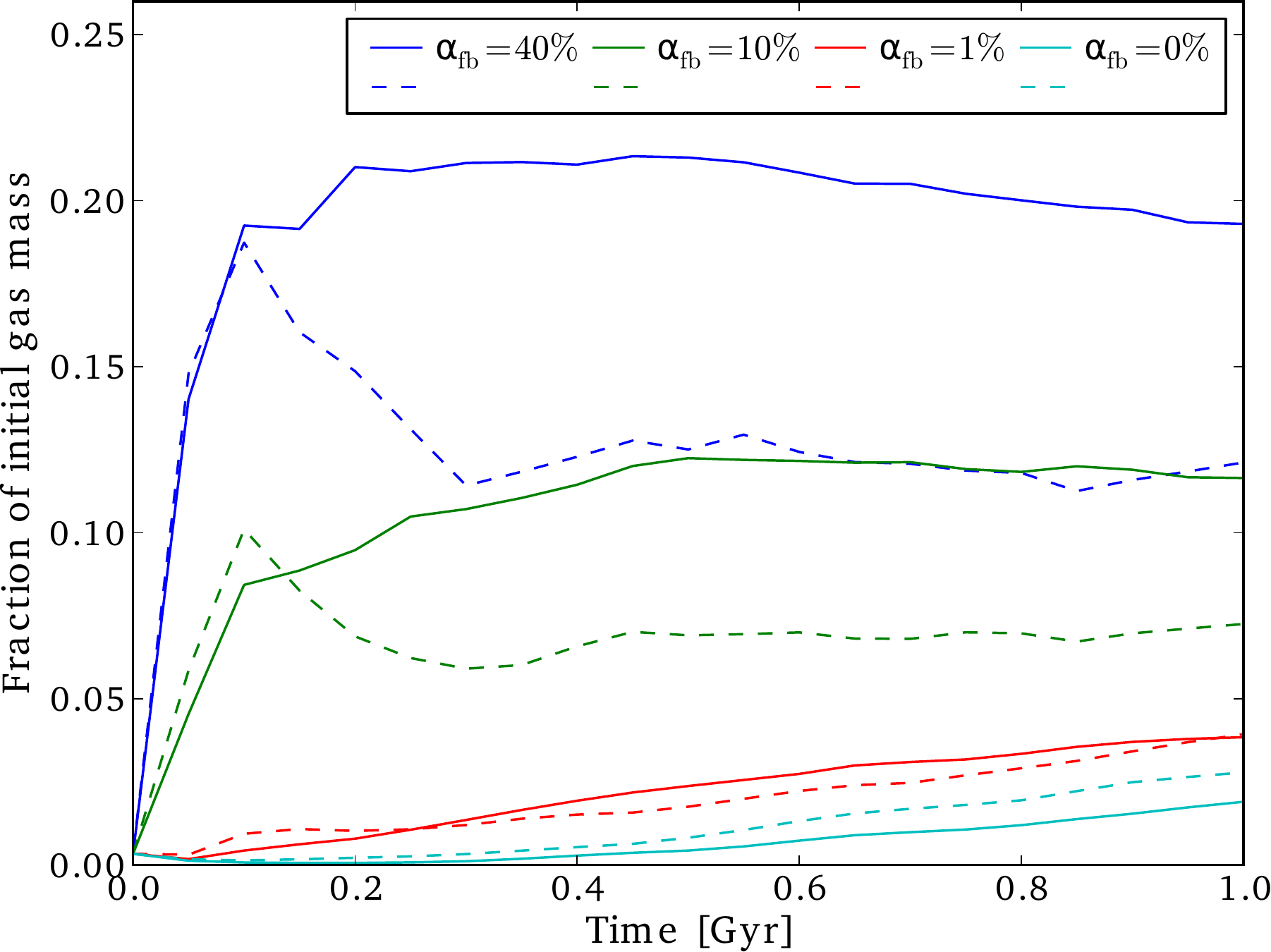}}
\caption{Fraction of gas beyond 1~kpc from the disc plane. Solid lines are runs with H$_2$ and dashed ones are runs without H$_2$.}
\label{outflow-fig}
\end{figure}

\begin{figure}[!h]
\centering
\resizebox{\hsize}{!}{\includegraphics{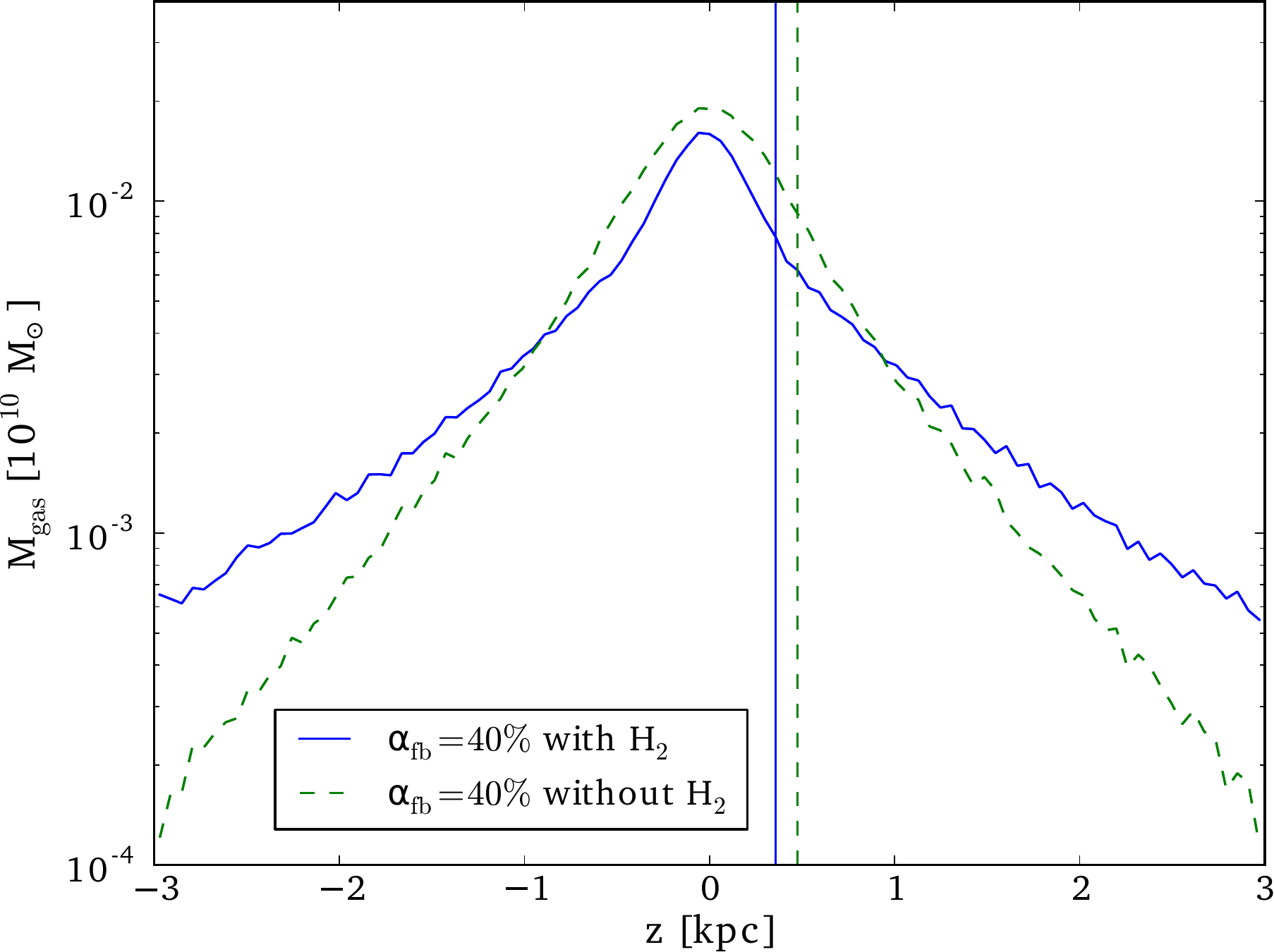}}
\caption{Vertical mass profile of the disc for radii R $>$ 15~kpc for runs with  $\alpha_{\mathrm{fb}} = 40 \%$. Solid line: run with H$_2$. Dashed line: run without H$_2$. The two vertical lines mark the characteristic vertical heights z$_{1/2}$.}
\label{rhoz-fig}
\end{figure}

The inclusion of molecular hydrogen has a significant impact on the vertical structure of the disc, first because 
the cold and dense gas concentrates in the disc middle plane, and second because of the impact of gas clumping 
on the distribution of feedback energy. Christensen et al. (2012a,b) include non-equilibrium formation of H$_2$, 
self-shielding and dust shielding of both HI and H$_2$ in galaxies extracted from cosmological simulations and 
explore the influence of including H$_2$ formation, for a fixed feedback efficiency. Similarly to their results,
we find that the introduction of H$_2$ makes the outer parts of discs denser while allowing the gas to go further away from the disc plane. This is because there is more star formation in the outer regions and therefore more feedback. As the volume density is also higher at large radii than with no molecular hydrogen, star formation takes place in denser regions, which increases the effect of feedback. In our feedback scheme, particles get velocity kicks weighted by the SPH kernel, so that the kicks are larger for particles closer to the new stellar particles. Feedback regulates star formation, but the SFR is still higher when H$_2$ can be formed and make the gas denser. We have performed simulations with various feedback efficiencies, helping to check this effect. Figure~\ref{outflow-fig} shows the fraction of gas that is further than 1~kpc from the disc (or radially further than 60~kpc from the centre of the galaxy): without feedback, the gas is only gravitationally heated and the effect of denser gas and higher clumping factor due to H$_2$ makes the fraction of gas leaving the disc smaller than for a purely atomic hydrogen gas. However, with feedback, 
there is a higher fraction of gas outside the disc when H$_2$ is included. This is due to the higher concentration of the 
gas that makes the feedback more efficient. In our simulations, the effect is especially striking in the outer parts of discs 
when H$_2$ formation is taken into account, as the vertical restoring force is lower there. 

We checked that the vertical density distribution is consistent with both a more concentrated disc and 
a higher fraction of gas leaving the disc. Figure~\ref{rhoz-fig} shows the vertical mass profile of the gas 
at large radii for $\alpha_{\mathrm{fb}} = 40 \%$: the gas is more concentrated in the disc plane, but the distribution 
has higher density ``tails'' when more efficient feedback in denser regions allows the gas to be expelled from the disc. 
Figure~\ref{rhoz-fig} shows the characteristic height z$_{1/2}$ of the gas, the distance from the disc plane for which 
the density equals half of the central density, as a function of radius for the various simulations. Especially at large 
radii, the characteristic height is lower for simulations with H$_2$ for all feedbacks as H$_2$ concentrates the gas in 
the middle plane. The fraction of gas beyond this height however increases with the feedback efficiency at large radii 
when H$_2$ is taken into account, as gas is then efficiently sent away from the disc plane by feedback. Without feedback, some difference remains, probably due to gravitational heating produced by a higher clumping. We see indeed a slight difference in vertical velocity dispersion.  

\subsubsection{Gas density profile}

\begin{figure}[!h]
\centering
\resizebox{\hsize}{!}{\includegraphics{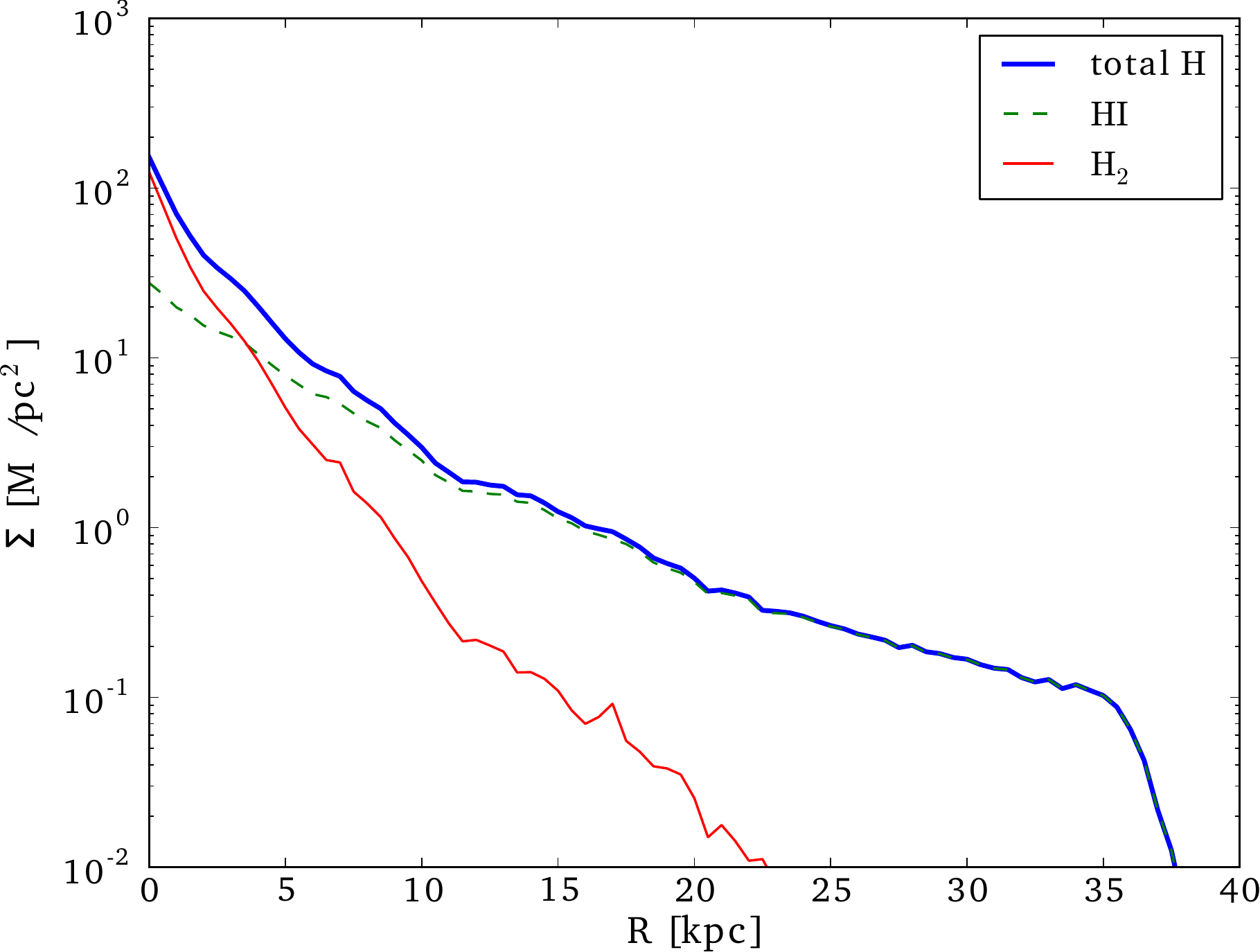}}
\caption{Radial distribution of the surface density of H$_2$, atomic gas HI and total hydrogen gas after 100~Myr for $\alpha_{\mathrm{fb}}=40 \%$ and a $\chi \times 500$ UV scaling factor.}
\label{lowrchar-fig}
\end{figure}

The galaxies we have considered until now have a rather low gas surface density. We also run simulations with smaller characteristic 
radii, and therefore higher surface densities. Figure~\ref{lowrchar-fig} displays the surface density for a Miyamoto-Nagai 
gas radius r$_g$ of 3.6~kpc. The transition radius between H$_2$ dominated and HI dominated regions, the radius at which $\Sigma_{\mathrm{H_2}}=\Sigma_{\mathrm{HI}}$, has then a value similar to the average observed by \citet{bigiel12} 
(their average value for nearby spiral galaxies observations is 14~M$_{\odot}$/pc$^2$). Previous galaxies belong to 
the lower surface density group observed by \citet{bigiel12}.
The effect of H$_2$ cooling is reduced in these galaxies with higher surface density, since a larger fraction
of the gas belong to the central regions, with much more efficient star formation than the outer parts. H$_2$ cooling 
is more important when there is more gas in the metal poor outer regions. 

\section{Conclusions} 
\label{conclu}

To explore the influence of molecular hydrogen in the physics of spiral discs, of star formation,
and gas reservoirs in galaxy evolution, 
we have implemented in Gadget-2 detailed cooling by metals, for temperatures as low as 100~K, 
and cooling by H$_2$ due to collisions with H, He and other H$_2$ molecules. 
The determination of the H$_2$ density is inspired by the KMT recipe  \citep{krum11}, 
using the stellar UV flux from young stars, and we study the influence of cold and dense 
molecular phase together with stellar feedback on star formation. This simple method requires some calibration when the resolution is changed. Using the instantaneous UV flux from young stars also means that some calibration must be done depending on the average SFR.
We have also implemented a stochastic star formation and a kinetic supernovae feedback whose efficiency was varied 
in simulations including cooling by atomic/ionised gas and/or molecular hydrogen. 
The evolution of the ISM of a galaxy depends on the parameters chosen for star formation and feedback. While we have focused on a star formation efficiency per free-fall time $c_*$ and a minimum density for star formation $n_{\rm H_{min}}$, changing these parameters can give a more heterogeneous ISM, but the resolution required to attempt to resolve the Jeans mass is more important.
The influence of H$_2$ in the formation of dense gas and star formation is very important in the
outer extended disk. 
Molecular hydrogen influences the vertical structure of the discs, especially when there is some stellar feedback: 
first H$_2$ makes the gas more concentrated in the middle layer of the disc plane, but second
the gas is also more susceptible of being ejected far from the disc, due to the higher efficiency of feedback 
in high density regions.   
 Correlating SFR and gas surface density, it is found that molecular gas is a much better tracer
of star formation than atomic gas, as is also observed in nearby galaxies.
We find that including 
molecular hydrogen allows some slow star formation to occur in the low metallicity outer parts of galaxies. 
If gas is accreted by the discs, it may help store some cold gas with a slow star formation.

\begin{acknowledgements}
We thank Yves Revaz, Martin Stringer, Maxime Bois and Benjamin L'Huillier for stimulating discussions. Computations have been done partly on the GENCI/TGCC machines and partly on the cluster funded by the European Research Council under the Advanced Grant Program 267399-Momentum.
\end{acknowledgements}

\bibliographystyle{aa} 

\bibliography{bibpaper1}

\end{document}